\documentclass[fleqn,usenatbib]{mnras}

\usepackage{newtxtext,newtxmath}
\usepackage{subfig}
\usepackage{pdflscape}
\usepackage[T1]{fontenc}
\newcommand{\angstrom}{\text{\normalfont\AA}}
\usepackage{multirow}
\usepackage{gensymb}
\DeclareRobustCommand{\VAN}[3]{#2}
\let\VANthebibliography\thebibliography
\def\thebibliography{\DeclareRobustCommand{\VAN}[3]{##3}\VANthebibliography}
\usepackage{hyperref}
\newcommand{\jwst}{{\textit{JWST}}}

\newcommand{\rasex}[3]{#1$^\mathrm{h}$\,#2$^\mathrm{m}$\,#3$^\mathrm{s}$}
\newcommand{\decsex}[3]
{+#1$^\circ$\,#2$^\prime$\,#3$^{\prime\prime}$}

\usepackage{graphicx}	
\usepackage{amsmath}	

\title[WEAVE first light: Stephan’s Quintet]{WEAVE First Light Observations:  Origin and Dynamics of the Shock Front in Stephan’s Quintet}

\author[M. I. Arnaudova et al.]{M. I. Arnaudova$^{1}$\thanks{E-mail: m.i.arnaudova@gmail.com},
S. Das$^{1}$, D. J. B. Smith$^{1}$, M. J. Hardcastle$^{1}$, N. Hatch$^{2}$, S. C. Trager$^{3}$, R. J. Smith$^{4}$, 
\newauthor A. B. Drake$^{1}$, J. C. McGarry$^{1}$, S. Shenoy$^{1}$, J. P. Stott$^{5}$, J. H. Knapen$^{6,7}$, K. M. Hess$^{8,9,10}$, K. J. Duncan$^{11}$, 
\newauthor A. Gloudemans$^{12}$, P. N. Best$^{11}$, R. Garc\'{i}a-Benito$^{10}$, R. Kondapally$^{11}$, M. Balcells$^{13,6,7}$, G. S. Couto$^{14}$, 
\newauthor D. C. Abrams$^{13}$, D. Aguado$^{6,7}$, J. A. L. Aguerri$^{6,7}$, R. Barrena$^{6,7}$, C. R. Benn$^{13}$, T. Bensby$^{15}$, 
\newauthor S. R. Berlanas$^{6,7}$, D. Bettoni$^{16}$,  D. Cano-Infantes$^{13}$, R. Carrera$^{17}$, P. J. Concepci\'{o}n$^{13}$, G. B. Dalton$^{18, 35}$, 
\newauthor G. D'Ago$^{19}$, K. Dee$^{13}$, L. Dom\'{i}nguez-Palmero$^{13,6,7}$, J. E. Drew$^{20}$, E. L. Escott$^{4}$, C. Fari\~{n}a$^{13, 6, 7}$, 
\newauthor M. Fossati$^{21,22}$, M. Fumagalli$^{23,24}$, E. Gafton$^{13}$, F. J. Gribbin$^{13}$, S. Hughes$^{19}$, A. Iovino$^{22}$, S. Jin$^{3}$,
\newauthor  I. J. Lewis$^{18}$, M. Longhetti$^{22}$, J. M\'{e}ndez-Abreu$^{6,7}$, A. Mercurio$^{25,26,27}$, A. Molaeinezhad$^{19}$, 
\newauthor E. Molinari$^{22}$, M. Mongui\'{o}$^{28,29}$, D. N. A. Murphy$^{19}$, S. Pic\'{o}$^{13}$, M. M. Pieri$^{30}$, A. W. Ridings$^{13}$, 
\newauthor M. Romero-G\'{o}mez$^{28,31}$, E. Schallig$^{18}$, T. W. Shimwell$^{32,9}$, R. Skvar\^{c}$^{13}$, R. Stuik$^{33,32}$, 
\newauthor A. Vallenari$^{16}$, J. M. van der Hulst$^{3}$, N. A. Walton$^{19}$, C. C. Worley$^{19,34}$\vspace{0.4cm}\\\\ 
 \textit{Affiliations are listed in Appendix \ref{sec:affiliations}}.
}

\date{Accepted XXX. Received YYY; in original form ZZZ}

\pubyear{2024}

\begin{document}
\label{firstpage}
\pagerange{\pageref{firstpage}--\pageref{lastpage}}
\maketitle

\begin{abstract}

We present a detailed study of the large-scale shock front in Stephan’s Quintet, a byproduct of past and ongoing interactions. Using integral-field spectroscopy from the new William Herschel Telescope Enhanced Area Velocity Explorer (WEAVE), recent 144 MHz observations from the LOFAR Two-metre Sky Survey (LoTSS), and archival data from the Very Large Array and \textit{James Webb Space Telescope} (\jwst ), we obtain new measurements of key shock properties and determine its impact on the system. Harnessing the WEAVE large integral field unit’s (LIFU) field of view (90 $\times$ 78 arcsec$^{2}$), spectral resolution ($R\sim2500$) and continuous wavelength coverage across the optical band, we perform robust emission line modeling and dynamically locate the shock within the multi-phase intergalactic medium (IGM) with higher precision than previously possible. The shocking of the cold gas phase is hypersonic, and comparisons with shock models show that it can readily account for the observed emission line ratios. In contrast, we demonstrate that the shock is relatively weak in the hot plasma visible in X-rays (with Mach number of $\mathcal{M}\sim2 - 4$), making it inefficient at producing the relativistic particles needed to explain the observed synchrotron emission. Instead, we propose that it has led to an adiabatic compression of the medium, which has increased the radio luminosity ten-fold. Comparison of the Balmer line-derived extinction map with the molecular gas and hot dust observed with JWST suggests that pre-existing dust may have survived the collision, allowing the condensation of H$_2$ - a key channel for dissipating the shock energy.

\end{abstract}




\section{Introduction}

Galaxy mergers and interactions are a cornerstone of our understanding of galaxy formation and evolution. They are 
believed to play key roles, not only in the mass build-up and morphological transformation of galaxies, but also for triggering mild and extreme starbursts and possibly active galactic nuclei \citep[AGN; e.g.][]{barnes1996transformations,hopkins2006unified,smith2010,satyapal2014galaxy,knapen2015interacting,cibinel2019early,diaz-garcia2020gas,pierce2023}. Mergers and interactions occur predominantly in  systems with high galaxy density and low velocity dispersion, making nearby compact galaxy groups ideal laboratories for studying their impact (e.g. \citealt{hickson1992dynamical}).

\begin{figure}
\centering
\includegraphics[width=1\columnwidth]{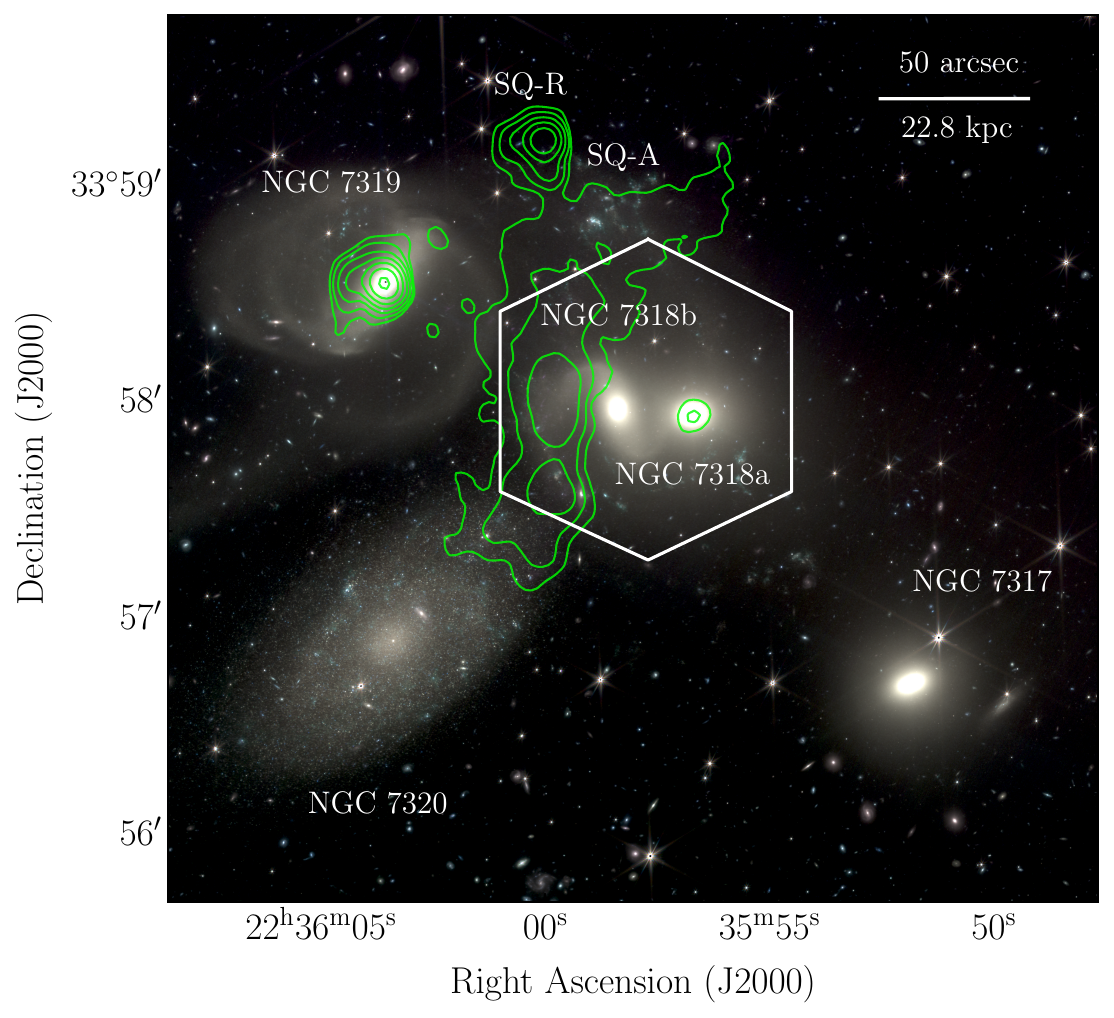}
    \caption{A composite image of Stephan's Quintet made with \jwst~NIRCam's F277W, F356W and F444W bands, orientated such that North is up, and East to the left. NGC 7320c is outside the FoV, located to the east of NGC 7319. The green contours represent the 144\,MHz radio flux density from LoTSS, with the lowest contour corresponding to a flux density of 1.2\,mJy beam$^{-1}$ and each subsequent contour increasing by a factor of two. The hexagon denotes the approximate coverage of the new WEAVE observations of the system.}
    \label{fig:jwst}
\end{figure}

One such system is Stephan's Quintet (hereafter SQ), which has been extensively studied across the electromagnetic spectrum ever since its discovery by \citet{stephan1877nebuleuses}.
SQ has a systemic redshift of $z=0.0215$ \citep{hickson1992dynamical}, and consists of three galaxies at its core (NGC 7317, NGC 7318a and NGC 7319 as shown in Figure \ref{fig:jwst}), which, according to the dynamical scenario proposed by \cite{moles1997dynamical} have experienced several interactions under the gravitational influence of a fourth galaxy with a similar redshift, NGC 7320c (the `old intruder', to the east of NGC\,7319, not shown in Figure \ref{fig:jwst}). A fifth nearby galaxy, NGC 7320, is considered to be a foreground source due to its discordant redshift \citep{burbidge1961further}. Unlike NGC 7317 and NGC 7318a, which have elliptical morphologies, NGC 7319 and the old intruder exhibit spiral structures. However, the process of interaction has stripped almost all of their interstellar medium \citep[ISM;][]{sulentic2001multiwavelength} into the intergalactic medium (IGM). The system is currently undergoing a new interaction between the IGM and NGC\,7318b, the `new intruder', as it enters the group for the first time at $\sim1000$ km s$^{-1}$ towards our line of sight. This has led to the formation of the large-scale shock region (hereafter LSSR), a giant filament of shocked gas, $\sim45$\,kpc in extent. The LSSR is located to the east of NGC 7318b, and has been studied extensively using X-rays (\citealt{trinchieri2005stephan}; \citealt{o2009chandra}), ionized gas emission (\citealt{xu2003physical}; \citealt{iglesias2012new}; \citealt{konstantopoulos2014shocks}; \citealt{rodriguez2014study}; \citealt{puertas2019searching}, \citeyear{puertas2021searching}), molecular H$_{2}$  gas (\citealt{appleton2006powerful}; \citealt{cluver2010powerful}; \citealt{appleton2017}; \citealt{appleton2023multiphase}) and at radio frequencies (\citealt{allen1972radio}; \citealt{vanderhulst1981VLA}; \citealt{williams2002vla}; \citealt{xu2003physical}; \citealt{nikiel2013intergalactic}).

Some of the key results from these works can be summarised as follows: X-ray observations of the LSSR by \citet{o2009chandra} revealed the presence of a hot plasma associated with a `ridge' with temperature $\sim 0.6\,$keV and metallicity $Z = 0.3 Z_{\odot}$, embedded within a larger hotter halo with similar metallicity, likely heated by previous dynamical encounters within the system. The hot gas in the ridge is believed to originate from an oblique shock heating a pre-existing \textsc{Hi} filament, presumably a remnant of a previous interaction (e.g \citealt{williams2002vla}; \citealt{jones2023disturbed}). Mid-infrared studies have found that the LSSR is further characterized by dominant warm H$_{2}$ emission, whose luminosity exceeds that of the hot X-ray plasma by a factor of 3 \citep{cluver2010powerful}, indicating that molecular hydrogen lines play a far greater role in the cooling of the gas than the thermal X-ray emission. By combining \textit{Spitzer Space Telescope} and \textit{James Webb Space Telescope} Mid-Infrared Instrument (hereafter \jwst\ MIRI) imaging, with CO (2-1) imaging spectroscopy from the Atacama Large Millimeter/submillimeter Array (ALMA), \cite{appleton2023multiphase} propose that this distinctive emission originates from the fragmentation of cold molecular clouds as they are mixed with post-shocked gas. These clouds were formed during the collision event and were subsequently subjected to ram-pressure stripping from the hot X-ray plasma generated by the shock, leading to the formation of smaller fog clouds emitting infrared H$_{2}$ radiation. Building upon the complexity of the system, \cite{guillard2022} used UV spectroscopic observations with the \textit{Hubble Space Telescope} Cosmic Origins Spectrograph (COS; \citealt{green2012COS}) and detected extremely broad Ly$\alpha$ emission (FWHM\,$\approx$\,2000\,kms$^{-1}$), whose radiated energy was found to be comparable to the much cooler H$_{2}$ and much hotter X-rays (within a factor of a few). These findings strongly support the picture that the galaxy-wide collision is driving a turbulent cascade of energy affecting all gas phases on both large and small scales, as initially proposed by \cite{guillard2009}. Finally, radio studies have revealed a steep non-thermal spectral index along the shock front between 1.43 -- 4.86\,GHz (observed with the Very Large Array; hereafter VLA; \citealt{thompson1980very}) and 4.85 -- 8.35\,GHz (using the Effelsberg 100-m Radio Telescope; \citealt{wielebinski2011effelsberg}), indicating the presence of old plasma. This suggests that the shock has been active for a significant period, allowing electrons to undergo various stages of interactions. 

These works have been complemented by optical IFU observations, which have transformed our ability to study the system by enabling us to trace its resolved kinematics and ionized gas properties. \cite{iglesias2012new} presented the first IFU observation of the LSSR in SQ using the Calar Alto 3.5\,m Telescope with the Potsdam Multi-Aperture Spectro-photometer (PMAS, \citealt{roth2005pmas}) to obtain pointings with a 16 $\times$ 16 arcsec$^{2}$ field of view (FoV) in three separate regions of the shock with an effective spectral resolution of $\mathrm{FWHM} = 3.6$\,$\angstrom$ across the wavelength coverage of $3810<\lambda<6809$\,\AA. The authors revealed the presence of three kinematic regimes: \textsc{Hii} regions with recession velocities consistent with the new intruder (5400 -- 6000 km\,s$^{-1}$), low-velocity shock-ionized gas with solar metallicity (5800 -- 6300\,km\,s$^{-1}$) and high-velocity shock-ionized gas with sub-solar metallicity ($\sim$6600\,km\,s$^{-1}$). 
These findings were confirmed and complemented by \cite{rodriguez2014study}, who used the same instrument but in PMAS fiber PAcK mode \citep[PPAK;][]{kelz2006pmas}, which traded lower spectral resolution (FWHM = 10\,$\angstrom$) for a larger FoV (74 $\times$ 65 arcsec$^{2}$) and extended wavelength coverage (3700 -- 7100 $\angstrom$). In addition, they have demonstrated by comparing with theoretical models that the nebular emission in SQ is consistent with a shock origin, with low density pre-existing gas, and shock velocities between $200-400$\,km\,s$^{-1}$. 

More recently, \cite{puertas2019searching,puertas2021searching} observed the entire system, making use of the large 11 $\times$ 11 arcmin$^{2}$ FoV of the SITELLE instrument \citep{grandmont2012final}, a Fourier transform spectrometer on the Canada France Hawaii Telescope. With three filters (SN1, SN2 and SN3) covering emission lines from [{O\,\sc ii}]\,3727\,$\angstrom$ to [{S\,\sc ii}]\,6716, 6731\,$\angstrom$ with a mean spectral resolution of R$\sim$500, R$\sim$760 and R$\sim$1560, respectively, the authors were able to study the star formation rates (SFR), oxygen and nitrogen-to-oxygen abundances of 175 {H${\alpha}$} regions in SQ. They found that the majority of star formation is present in the starburst region A (SQ-A, the region to the North end of the shock, which appears blue as seen in Figure \ref{fig:jwst} as a result of recent star formation), suggesting that prior to the collision there was little ongoing star formation in the IGM. Their work also revealed the presence of two chemically different regions, with the low/high-metallicity material primarily associated with a relatively low/high radial velocity, suggesting that the metal-rich component comes from the inner part of the new intruder.
 
In this study, we build upon these results, by combining the first-light data from the William Herschel Telescope Enhanced Area Velocity Explorer large integral field unit (WEAVE-LIFU; \citealt{dalton2012weave}) with a range of new and archival multi-wavelength data, including 144\,MHz radio observations from the second data release of the LOFAR Two-metre Sky survey (LoTSS; \citealt{shimwell2022lofar}). The combination of the arcminute-scale FoV (90 $\times$ 78 arcsec$^{2}$) of the WEAVE-LIFU, 2.6\,arcsec spaxels, high spectral resolution ($R \sim 2500$), and virtually-complete wavelength coverage across the optical wavelengths ($3660-9590$\AA) sets the new data apart from those obtained in the past (e.g. WEAVE has a larger FoV than PMAS, plus higher spectral resolution than in any pre-existing data, coupled with the large wavelength coverage). These features allow us to study the dynamics and other properties of the shock front in greater detail than previously possible. In addition, we expand the multi-wavelength picture of the system by considering low-frequency radio data, which traces older, more diffuse emission, allowing us to gain new insights into the complex history of Stephan's Quintet. 

The structure of this paper is as follows: Section \ref{sec:data} describes the data used in this work, including WEAVE-LIFU and LoTSS observations along with additional auxiliary data from \jwst~and the VLA. Section \ref{sec:method} gives an overview of the spectral fitting process used in this work. Section \ref{sec:results} presents an overall view of SQ based on the multi-wavelength data we have assembled, while Section \ref{sec:shock properties} investigates the shock region in more detail. Finally, Section \ref{sec:conclusions} summarises and discusses our findings. 

For consistency with previous works (e.g. \citealt{appleton2017}; \citealt{guillard2022}; \citealt{appleton2023multiphase}), we assume a distance of 94\,Mpc to Stephan’s Quintet (for $H_0=70\,$km\,s$^{-1}$\,Mpc$^{-1}$ and a heliocentric group velocity of 6600\,km\,s$^{-1}$). At this distance, $1\arcsec$ corresponds to a linear scale of 0.456\,kpc.

\section{Data}\label{sec:data}

\subsection{WEAVE First Light Data}\label{sec:data_weave}

LIFU observations of Stephan's Quintet were taken on 25 October 2022, and released to members of the WEAVE consortium on 12 December 2022 as part of WEAVE's first light data release. The WEAVE LIFU consists of 547 closely-packed optical fibres with a filling factor of 0.55, each 2.6 arcsec in diameter, together providing a FoV of 90 $\times$ 78 arcsec$^{2}$. The data were acquired in two observing modes -- `high-resolution', which offers a spectral resolution of $R\sim 10,000$ across the wavelength range of $4040<\lambda<6850$\,\AA, and `low-resolution', which has $R \sim 2500$ across $3660 < \lambda < 9590$\,\AA. In this work, we focus on the `low-resolution' mode data due to its higher signal-to-noise ratio, which enables us to study the emission lines in the LSSR on a per-spaxel basis.

The observations were taken at a mean airmass of 1.01, in 1.67\,arcsec seeing, and at a position angle of 120$^\circ$ centred near $(\alpha, \delta) = ($\rasex{22}{35}{59.3}, \decsex{33}{58}{12.3}$)$ between the galaxies NGC\,7318a and NGC\,7318b (see Figure \ref{fig:jwst} for details, in which the hexagonal WEAVE FoV is outlined in white). The observations consisted of six exposures, each 1020 seconds in duration, using both the blue ($3660-6060$\AA) and red arms ($5790-9590$\AA) with the native $1\times$ spectral binning giving a scale of  0.30 and 0.48\,$\angstrom$\,pixel$^{-1}$, respectively. Between exposures, the telescope pointing was adjusted using the default six-point dither pattern, designed to permit complete spatial sampling once the individual observations are combined. 
The resulting data were fully reduced using the Core Processing System \citep[CPS;][]{Walton2014} WEAVE pipeline located at the Cambridge Astronomical Survey Unit (CASU).
More information about the data reduction process is given by \citet{Jin2023}, however, the main steps can be summarized as follows. The data are bias- and flat-field corrected, and observations of a quartz-halogen lamp are used to trace out the location of each LIFU fibre's spectrum along the CCDs, to enable the spectra to be extracted. A wavelength solution is produced based on observations of ThArCr arcs, and a flux scale derived based on observations of white dwarfs. `Superstack' data cubes for each arm are then produced by reconstructing the 3D spectra onto a fixed wavelength grid with 0.5\,$\angstrom$ pixels. Spectra are produced at each location using the six individual exposures, conserving flux and inverse variance, and accounting for the dithering pattern. Finally, sky subtraction is performed using sky spectra obtained simultaneously from dedicated sky fibres positioned outside the main FoV. The sky spectrum is modelled using a principal component analysis algorithm that accounts for source contamination in the sky fibres, which will be described in detail by a future work.

The resulting data cubes are stored in FITS format, are $178 \times 188$ 0.5\,arcsec spaxels on a side, and consist of seven extensions: (i) data, (ii) inverse variance, (iii) data without sky-subtraction, (iv) inverse variance without sky-subtraction, (v) sensitivity function used for flux calibration, (vi) white-light image collapsed in the wavelength direction, (vii) white-light inverse variance. This work uses version 1 of the fully reduced data, which can be downloaded from the WEAVE Archive System (WAS\footnote{\url{http://portal.was.tng.iac.es}}).

\subsection{Radio Observations}
The 144\,MHz observations of SQ are included in the second data release of the LOFAR Two-metre Sky survey (LoTSS DR2) and are accessible on the LOFAR surveys website\footnote{\url{https://lofar-surveys.org/dr2\_release.html}}. A comprehensive description of the data processing is given by \cite{shimwell2022lofar}. To briefly summarise, the 144\,MHz data were reduced using the fully automated LoTSS processing pipeline, which corrects for direction-independent instrumental effects, as well as ionospheric distortions that vary with time and direction. 
The resulting images have a resolution of 6 arcsec, with an RMS uncertainty of $\sim$130\,$\mu$Jy\,beam$^{-1}$ near SQ. Due to the excellent $uv$ coverage provided by the Dutch baselines, the LoTSS images are sensitive to emission on all scales up to $\sim 1^\circ$ including compact sources, as well as those with diffuse, extended emission (such as SQ).

Additionally, we used publicly available observations of SQ using the Karl G. Jansky Very Large Array \citep[VLA;][]{perley2011vla}. Project ID 18B-080 (PI: Blazej Nikiel-Wroczynski) observed the system at 1.7\,GHz in C-array configuration, while project ID AS939 (PI: M. Soida) made an observation at 4.86\,GHz in the D-array configuration. These datasets are selected for this study because of their high sensitivity to large-scale extended emission (6 arcmin at 4.86\,GHz and 17 at 1.7\,GHz) and their angular resolution, which is similar to that of the LoTSS data (see Table \ref{tab:rad_images} for details). We used the Common Astronomy Software Applications \citep[\texttt{\textsc{CASA}};][]{Bean2022CASA} VLA calibration pipeline to reduce the 1.7\,GHz dataset, while the 4.86\,GHz dataset was processed using the \texttt{\textsc{VLARUN}} routine within the Astronomical Image Processing System \citep[\texttt{\textsc{AIPS}};][]{Greisen2003aips}. In both cases, we conducted multiple iterations of phase-only and phase$+$amplitude self-calibration to remove residual errors.
More information about the radio images covering SQ can be found in Table \ref{tab:rad_images}.

\begin{table}
    \centering
    \caption{Information about the radio images of SQ used in this work, with details of the synthesized beam size, the peak flux densities and rms noise values.}
    \begin{tabular}{l|c|c|c}
        \hline
         Image  &  Beam FWHM & Peak flux& RMS  \\
                & arcsec$^{2}$ & $\mathrm{mJy}$ & $\mu\mathrm{Jy~beam}^{-1}$ \\
        \hline
         LoTSS 144\,MHz &  $6\times6$         &  662 &    129.6 \\
         VLA C-Array 1.7 GHz &  $13.35\times11.95$ &  62  & 16.3 \\
         VLA D-Array 4.86 GHz &  $13.63\times12.63$ &  3.5 & 8.6 \\
         \hline
    \end{tabular}
    \label{tab:rad_images}
\end{table}

\subsection{Additional Auxiliary Data}\label{sec:data_jwst} 

\jwst~NIRCam and MIRI observations of SQ are available as part of the Early Release Observations (ERO) described in \cite{Pontoppidan2022}. NIRCam data were obtained using broad-band filters such as F277W, F356W and F444W (with central wavelengths of 2.776$\mu$m, 3.566$\mu$m and 4.401$\mu$m, respectively) and the FULLBOX dither pattern to produce a 6.3$ \times $7.3 arcmin$^{2}$ rectangular mosaic covering all galaxy members, except NGC\,7320c. These data are well-suited for providing an overview of the system, effectively highlighting the complexity and intricate details of SQ, as shown in Figure \ref{fig:jwst}.

In addition, the MIRI data were obtained in three bands (F770W, F1000W and F1500W with central wavelengths 7.7\,$\mu$m, 10\,$\mu$m and 15\,$\mu$m, respectively) over a smaller area using four tiles centered on NGC 7318a/b, NGC 7319 and NGC\,7320c. The significance of these bands lies in their ability to capture specific spectral features. The F770W band is found to trace the polycyclic aromatic hydrocarbon (PAH) complex when star formation dominates, and the 0–0 S(5) H$_{2}$ emission otherwise. The F1000W band, on the other hand, is dominated by emission coming from the S(3) mid-IR pure rotational H$_{2}$ line, whereas the F1500W filter captures the faint dust continuum of the IGM (see \citealt{appleton2023multiphase} for further details). All these data are publicly available and were downloaded from the Mikulski Archive for Space Telescopes\footnote{\href{https://archive.stsci.edu/doi/resolve/resolve.html?doi=10.17909/dfsd-8n65}{doi:10.17909/dfsd-8n65}}.

\begin{figure*}
\centering
\includegraphics[width=0.95\textwidth]{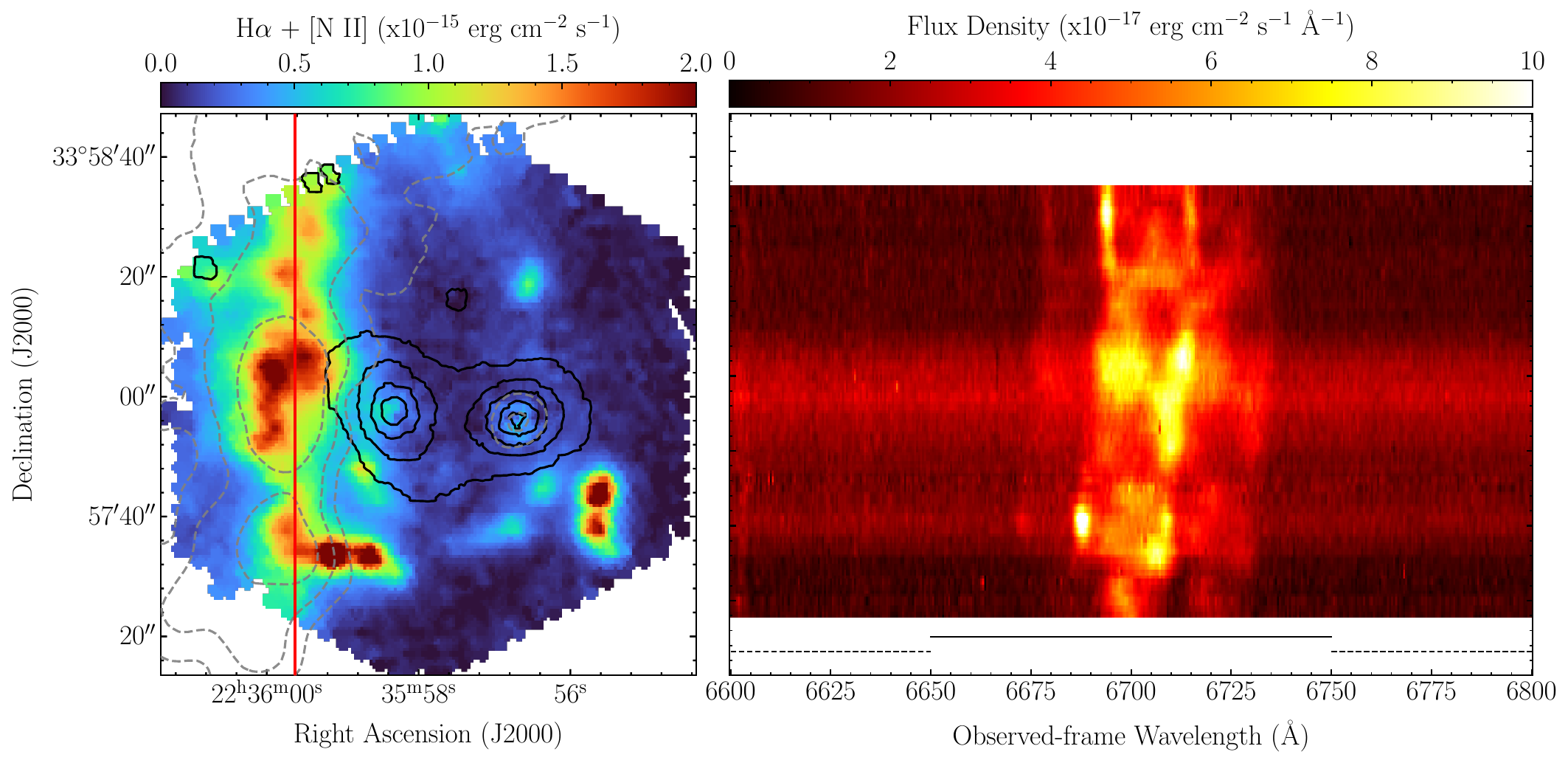}
    \caption{The complex velocity structure in the LSSR. The left panel presents the H$\alpha$ + [N\,\textsc{ii}] complex obtained by summing a spectral region  100 $\angstrom$ wide, centred on H$\alpha$, continuum-subtracted in each spaxel using a linear approximation (the regions used for emission lines and continuum estimation are indicated in the lower part of the right panel by the solid and dashed lines, respectively). The black contours indicate the location of NGC7318a\slash b in the white light image, constructed by collapsing the data cubes in the wavelength direction; each successive contour represents a factor of two increase in surface brightness. The grey dashed contours represent the LoTSS data, which are the same contours shown in green over a larger area in Figure \ref{fig:jwst}. The right panel shows the 2D spectrum (flux density in wavelength versus spatial position) in the vicinity of the H$\alpha$ + [N\, \textsc{ii}] complex across the region denoted by the red line in the left panel.}
    \label{fig:vel_map}
\end{figure*}

\section{Spectral Fitting Method}\label{sec:method}

To study the nebular gas properties of SQ, and the LSSR in particular, we fitted the principal emission line complexes (including the brightest species; H$\beta\,\lambda$4861, [\textsc{Oiii}]\,$\lambda\lambda$4959\slash 5007, H$\alpha$\,$\lambda$6563, and [\textsc{Nii}]\,$\lambda\lambda$6548,6583) in all available 0.5\,arcsec spaxels, using multiple Gaussian components \citep[in a manner similar to][]{comeron2021}. To determine the optimal number of components required for a given spaxel, we used the Bayesian Information Criterion (BIC), which enables us to find a balance between the goodness of fit and the number of free parameters in our model. Such a technique is necessary because of the complex velocity structure of the emission lines in SQ. This is evident in Figure \ref{fig:vel_map}, in which the left-hand panel shows the approximate integrated emission from the H$\alpha$ + [N\,\textsc{ii}] complex, while the right-hand panel shows a position-velocity diagram along the North-South direction of the peak emission (the location is indicated by the red vertical line in the left-hand panel). The H$\alpha$ + [\textsc{Nii}] emission has been calculated by summing over a spectral window 100\,$\angstrom$ wide centered on H$\alpha$ for each spaxel after subtracting the continuum (approximated using a linear model). Despite the simple continuum model (which is responsible for the faint emission apparently coincident with NGC7318a \& b; see Section \ref{sec:weave_overview} for details), we can see that the majority of the H$\alpha$ emission is concentrated in the region surrounding the LSSR, located to the East of the two central galaxies (NGC\,7318a \& b). The approximate location of the LSSR is indicated by the dashed contours in the left panel of Figure \ref{fig:vel_map} which has been derived on the basis of the 144\,MHz emission (see caption for details). Similarly, the locations of the two central galaxies in the WEAVE data cube are indicated by the solid contours in the same figure. 

The first step in our modelling approach is to correct the spectra for Galactic extinction (which is considerable, with a V-band extinction of $A_{V}\sim0.2$) in the observed frame by using the re-calibrated reddening data, $E(B-V)$, from \citet*{schlegel1998maps}, along with the Milky Way reddening curve from \cite{fitzpatrick1999correcting} for an extinction-to-reddening ratio of $R_{V}=3.1$. For each spaxel, we fitted spectral windows of interest from the blue and red arms simultaneously, modelling the local continuum near each line complex (i.e. covering from H$\beta$ to [\textsc{Oiii}]\,$\lambda$5007 and from [\textsc{Nii}]\,$\lambda$6548 to [\textsc{Nii}]\,$\lambda$6583, and later extending to [\textsc{Sii}]\,$\lambda6731$) with a straight line, and each emission line with up to four Gaussian components. Each spaxel was modelled independently of its neighbours. The simplistic, straight-line approach to continuum subtraction was adopted since our primary interest is in the regions away from the galaxies, where the continuum is faint and free from significant structure (see e.g. the right panel of Figure \ref{fig:vel_map}).

To reduce the number of free parameters in our model and obtain good constraints on the velocity structure of the system, we fixed the relative velocity offsets and line widths for each Gaussian components across all of the lines modelled. In addition, we set the flux ratios for the [\textsc{Oiii}]\,$\lambda\lambda$4959/5007 and [N\,\textsc{ii}]\,$\lambda\lambda$6548/6583 doublets to their predicted values (2.98 for [\textsc{Oiii}] following \citealt{dimitrijevic2007flux}, and 3.05 for [\textsc{Nii}] following \citealt{dojvcinovic2023flux}). As the number of free parameters is still considerable (e.g. 28 for modelling six emission lines with four Gaussian components), we employed the Markov Chain Monte Carlo (MCMC) method with 250 walkers running for 5000 steps. To determine reliable flux estimates and associated uncertainties for each of the lines modelled, we discarded the first 4500 steps from each walker's chain as burn-in. These results were visually inspected to ensure that the chains converged and that the results were robust.

Finally, to determine which of the four models to select (one for each number of Gaussian components), we used the minimal BIC value, which is computed in the following way, assuming that the model errors are independent and identically distributed according to a normal distribution:
\begin{equation}
    \mathrm{BIC} = k\ln(n) + \chi^{2}. 
\end{equation}
Here $n$ is the number of data points, $k$ is the number of free parameters and $\chi^{2}$ is the chi-squared parameter. In Appendix \ref{appendix2}, we provide examples of fits consisting of all four models identified at representative locations within SQ to demonstrate the performance of this method.

From the best-fit models for each emission line species, we obtain the total emission line flux, equivalent width (EW), line-of-sight velocity ($v$) and velocity dispersion ($\sigma$) including all of Gaussian components. While determining the line flux and the EW in the presence of multiple velocity components is straightforward, this makes obtaining representative velocities and velocity dispersions more complicated. For the line-of-sight velocity, we compute the weighted average velocity, where the weights are given by the emission line flux for each Gaussian component. Similarly, the velocity dispersion is calculated from the full width at half maximum (FWHM), which is determined by using the combined profile fit for each species, irrespective of the line shape. The uncertainties associated with each spectral property are evaluated by propagating the values extracted from the MCMC chains. For the entirety of the analysis, we concentrate on spaxels that have a $\ge 3\sigma$ detection in all four bright lines (i.e.  H$\beta$\,$\lambda$4861, [\textsc{Oiii}]\,$\lambda$5007, H$\alpha$\,$\lambda$6563, and [\textsc{Nii}]\,$\lambda$6583), since these are the spaxels with the best velocity constraints that enable us to confidently de-blend H$\alpha$ from the [\textsc{Nii}] line.

\section{A Multi-wavelength view of Stephan's Quintet}\label{sec:results}

\subsection[]{Stephan's Quintet as traced by H$\alpha$ with WEAVE}\label{sec:weave_overview}
\begin{figure*}
\centering
\includegraphics[width=.9\textwidth]{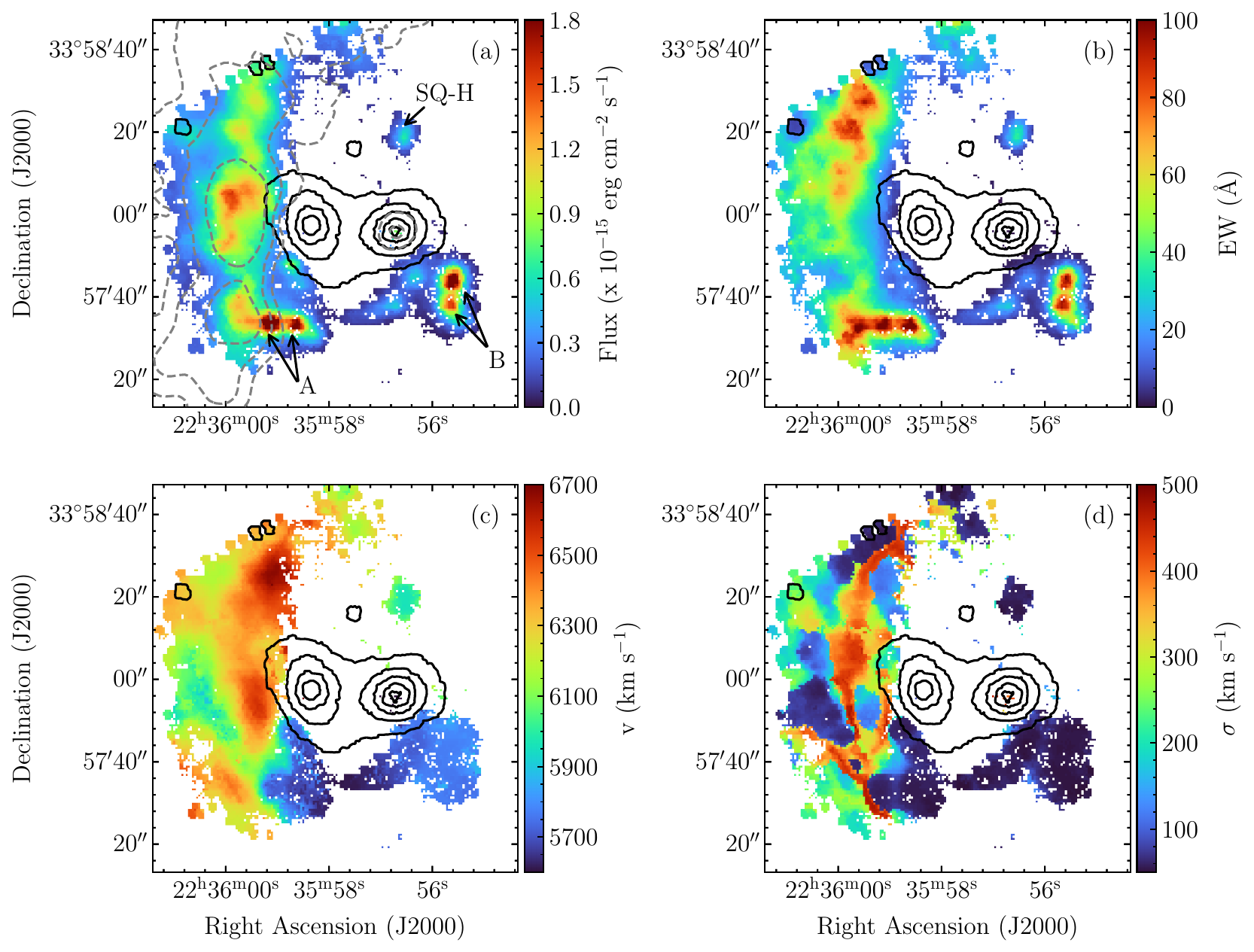}
    \caption{The fitted properties of the H$\alpha$ emission for all spaxels with a $3\sigma$ detection in each of the four bright lines (H$\beta$\,$\lambda$4861, [\textsc{Oiii}]\,$\lambda$5007, H$\alpha$\,$\lambda$6563, and [\textsc{Nii}]\,$\lambda$6583). 
    The panels, which are labelled in the upper-right corner, show: (a) line flux, (b) Equivalent Width, (c) line-of-sight velocity, and (d) velocity dispersion. In the upper-left panel (a) we have overlaid grey contours to indicate the LoTSS 144\,MHz flux density as in Figure \ref{fig:jwst}. 
    The black contours in each panel show the location of the central galaxies, based on the white light image, as in Figure \ref{fig:vel_map}. }
    \label{fig:Halpha}
\end{figure*}

\begin{figure*}
\centering
\includegraphics[width=1.\textwidth]{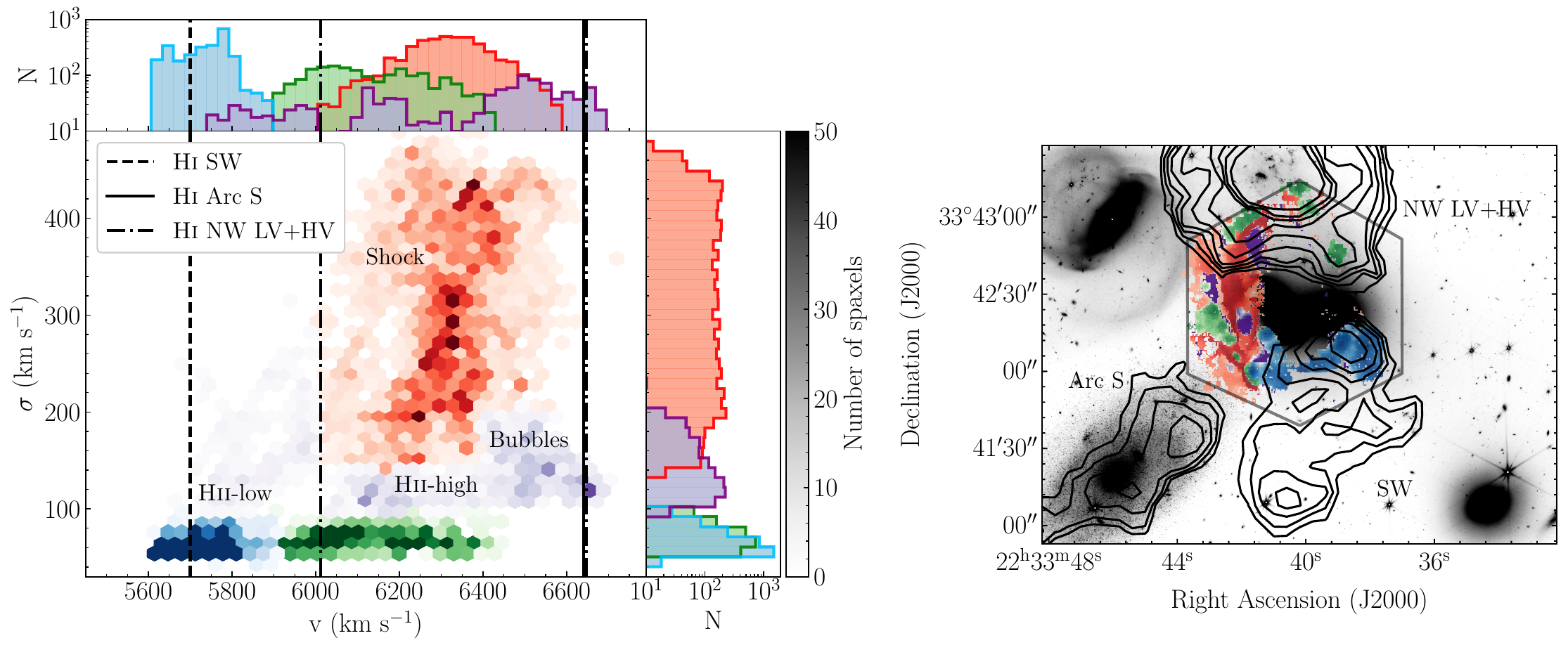}
    \caption{The line-of-sight versus velocity dispersion diagram used to dynamically identify regions of interests in SQ. The left panel shows the 2D distribution of the line-of-sight velocity and velocity dispersion for the different regions denoted, along with their 1D marginal distributions. The solid, dashed and dot-dashed lines represent the line-of-sight velocities of the $\textsc{Hi}$ filaments from \citet{williams2002vla}. The colourbar, applicable to all four colour sets, denotes the number of spaxels. The right panel shows the region around the WEAVE FoV (approximate location denoted by the black hexagon) with spaxels colour-coded on the basis of the features identified in the left panel, overlaid on the \jwst\ NIRCam F356W image in greyscale. For context, we have further added the $\textsc{Hi}$ column density map, based on reprocessing the data from \citeauthor{williams2002vla} (\citeyear{williams2002vla}; K. Hess, \textit{private communication}), by integrating over a velocity range of 5618$ < $V$ <\ $6776\,km\,s$^{-1}$. The contour levels are  7.5, 15, 23, 32, 44, 61, 87, 128 \&\ $180 \times 10^{19}$ atoms\,cm$^{-2}$, similar to Figure 5 of \citet{williams2002vla}, with the different $\textsc{Hi}$ filaments labelled.}
    \label{fig:v_sigma}
\end{figure*}

For a general overview of the system, in Figure \ref{fig:Halpha} we show the spectral properties of the H$\alpha$ emission in greater detail than previously possible. Figure \ref{fig:Halpha} contains four panels showing (a) the integrated H$\alpha$ flux, (b) the H$\alpha$ EW, (c) the weighted-average H$\alpha$ velocity, and (d) the H$\alpha$ velocity dispersion. As in Figure \ref{fig:vel_map}, black contours have been overlaid in each panel to indicate the positions of the central galaxies (NGC\,7318a \& b) based on the white light image. 
What is immediately remarkable upon even a cursory inspection of Figure \ref{fig:Halpha} is that the structures apparent in the figure show significant continuity between neighbouring spaxels, including on scales larger than the 2.6\arcsec LIFU fibres (which are themselves larger than the seeing at the time of the observations). Although the data themselves are not independent on the scale of the fibres (due to the correlated noise arising from the manner in which the cube has been reconstructed from the dithered integrations), the continuity extends to features approaching the arcmin scale of the FoV, and this is remarkable since each spaxel in the reconstructed WEAVE cube has been fit independently. This continuity\slash smoothness offers great encouragement that our fitting is performing in a robust manner. 

In panel (a) the integrated flux in the H$\alpha$ emission line -- after accounting for the [{N\,\sc ii}] contribution in the fitting -- does not reveal any H$\alpha$ emission associated with the cores of NGC7318a\slash b, in contrast to Figure \ref{fig:vel_map} where this results from a simple continuum-subtraction process in the presence of the bright stellar continuum at those locations (visual inspection confirms that all emission lines considered are absent in the cores of NGC\,7318a \& b as demonstrated in Appendix \ref{appendix2}). 
The H$\alpha$ flux map has been overlaid with grey dashed contours representing the LoTSS 144\,MHz flux density, and there is a clear spatial association between the H$\alpha$ intensity and the 144\,MHz flux density in the region surrounding the shock, suggesting a common excitation mechanism. In the South-Eastern and South-Western areas (labelled ``A'' and ``B'') however, there are distinct clusters of intense H$\alpha$ emission with no overlapping contours associated with 144\,MHz flux density. In panel (b) we can see large equivalent widths visible throughout the system, which offer some support for our choice to use a simple linear continuum model for our line fitting.

The velocity map in panel (c) shows complex velocity structure; similarly to previous works (e.g. \citealt{iglesias2012new}, \citealt{konstantopoulos2014shocks}, \citealt{puertas2019searching}), we can identify a  $v=$ 5600-5900\,km\,s$^{-1}$ component associated with the new intruder, as well as regions with higher velocity in the LSSR. However, unlike previous studies, WEAVE has the sensitivity and spatial coverage to provide a virtually complete sampling of the LSSR. While panels (b) and (c) reveal smoothly varying EW and line-of-sight velocities, the map of velocity dispersion shown in panel (d) reveals more complex structure\footnote{Recall that the velocity dispersion is computed using all velocity components of the fitted model, meaning that the large velocity dispersion could arise from one very broad component or from two or more narrower ones spread over a wider velocity range.}. Intriguingly, the region with the highest velocity dispersion ($\sigma \gtrsim 200$\,km\,s$^{-1}$) is spatially coincident with the peaks in the H$\alpha$ and 144\,MHz emission, which is surrounded by significantly lower $\sigma$ areas, including several `bubble'-like structures with $\sigma \sim 150\,$km\,s$^{-1}$. 

To put these features into context, in the left panel of Figure \ref{fig:v_sigma} we show a 2D histogram of the line-of-sight velocity $(v)$ and velocity dispersion $(\sigma)$ for the spaxels in the cube, which has been decomposed into three main features that we will define in the next paragraph, each shown in a particular colour. In the right panel, the spaxels in each velocity feature are located within the WEAVE FoV, and colour-coded in the same way. Although this decomposition is clearly subjective (and therefore open to some cross-contamination between regions), the features identified in velocity space appear contiguous in the projected FoV, and our conclusions are insensitive to the details of this decomposition (e.g. the precise locations of the regions, or whether the histogram is weighted by the H$\alpha$ flux density in each spaxel), suggesting that the regions we have identified are physically plausible.

The regions are identified as follows. We first identify two distinct regions with low velocity dispersion ($\sigma < 100$\,km\,s$^{-1}$), one with relatively low line-of-sight velocity ($v < 5900$\,km\,s$^{-1}$) that is consistent with the new intruder and coloured blue in Figure 4 (hereafter ``\textsc{Hii}-low''), and another located at higher line-of-sight velocity ($v=5900-6400$\,km\,s$^{-1}$) and coloured green in Figure 4 (hereafter ``\textsc{Hii}-high''). The properties of these regions (low velocity dispersion and high EW as shown in the top right panel of Figure \ref{fig:Halpha}) correspond to typical attributes of {H\,\sc{ii}} regions \citep[e.g.][]{zaragoza2015,Lima2020}. The velocity of \textsc{Hii}-low being consistent with the new intruder provides strong evidence of association, and is consistent with previous studies which suggested that these \textsc{Hii} regions have survived the collision of NGC\,7318b with SQ (e.g. \citealt{iglesias2012new}). Notable in \textsc{Hii}-high is the relatively isolated (in projection) and previously unknown SQ-H, labelled in the North-West corner of Figure \ref{fig:Halpha}\,(a). This fainter \textsc{Hii} region has a velocity between those of the LSSR and the new intruder, consistent with the velocities of \textsc{Hi} filaments described in \citet{williams2002vla}, indicating that it may be evidence of shock-induced star formation.

Finally, we identify a third region as having velocity dispersion $\sigma \gtrsim 150$\,km\,s$^{-1}$ and line-of-sight velocity $v = 6000-6600$ km\,s$^{-1}$, which we define hereafter as the shock region. This region is coloured red in Figure \ref{fig:v_sigma}, and we will refer to this dynamical definition of the shock repeatedly in the following sections. For completeness, we have also outlined the locations of the remaining spaxels that do not correspond to any of the three previously identified regions; we refer to these regions hereafter as the ``bubbles'' on account of their projected appearance (and they are shown in purple in Figure \ref{fig:v_sigma}). The line profiles associated with the bubbles (see Appendix \ref{appendix2} for an example) clearly differ from those of the neighbouring shocked region and \textsc{Hii}-high, underscoring the complicated velocity structure in SQ, which does not appear to be a line-of-sight effect (as demonstrated by the position-velocity diagrams in Appendix \ref{appendix2}). 

Overlaid on the left panel of Figure \ref{fig:v_sigma} are vertical lines indicating the velocities of the various \textsc{Hi} structures identified in \citet[]{williams2002vla}, and the imaging of these data (K. Hess, \textit{private communication}) is also overlaid as contours in the right panel of Figure \ref{fig:v_sigma}. The \textsc{Hi} data are particularly interesting in this context since there is a clear correspondence between the velocity of \textsc{Hii} low, the South-Western \textsc{Hi} cloud (\textsc{Hi} SW) and the new intruder, as well as between SQ-H and the North-Western \textsc{Hi} low-velocity component (\textsc{Hi} NW LV), and between the `bubbles', the Southern \textsc{Hi} arc (\textsc{Hi} Arc S), and the high-velocity component of the North-West \textsc{Hi} emission (\textsc{Hi} NW HV). Taken together, these results strongly suggest that the collision of NGC7318b with the IGM has ionized the cool gas scattered throughout the region by the previous galaxy interactions, perhaps explaining the deficiency of \textsc{Hi} that is apparent near the LSSR (though of course galaxy interactions rarely result in homogeneous distributions of stripped gas, at least until the system has had time to relax). This finding has been previously suggested (e.g. \citealt{iglesias2012new}), however, not with such a detailed kinematic decomposition.

\subsection{Extinction, dust and gas with WEAVE and \jwst\ MIRI}\label{sec:dust properties}

\begin{figure*}
    \centering
    \includegraphics[trim={0cm 1cm 3cm 0cm},clip,width=0.8\textwidth]{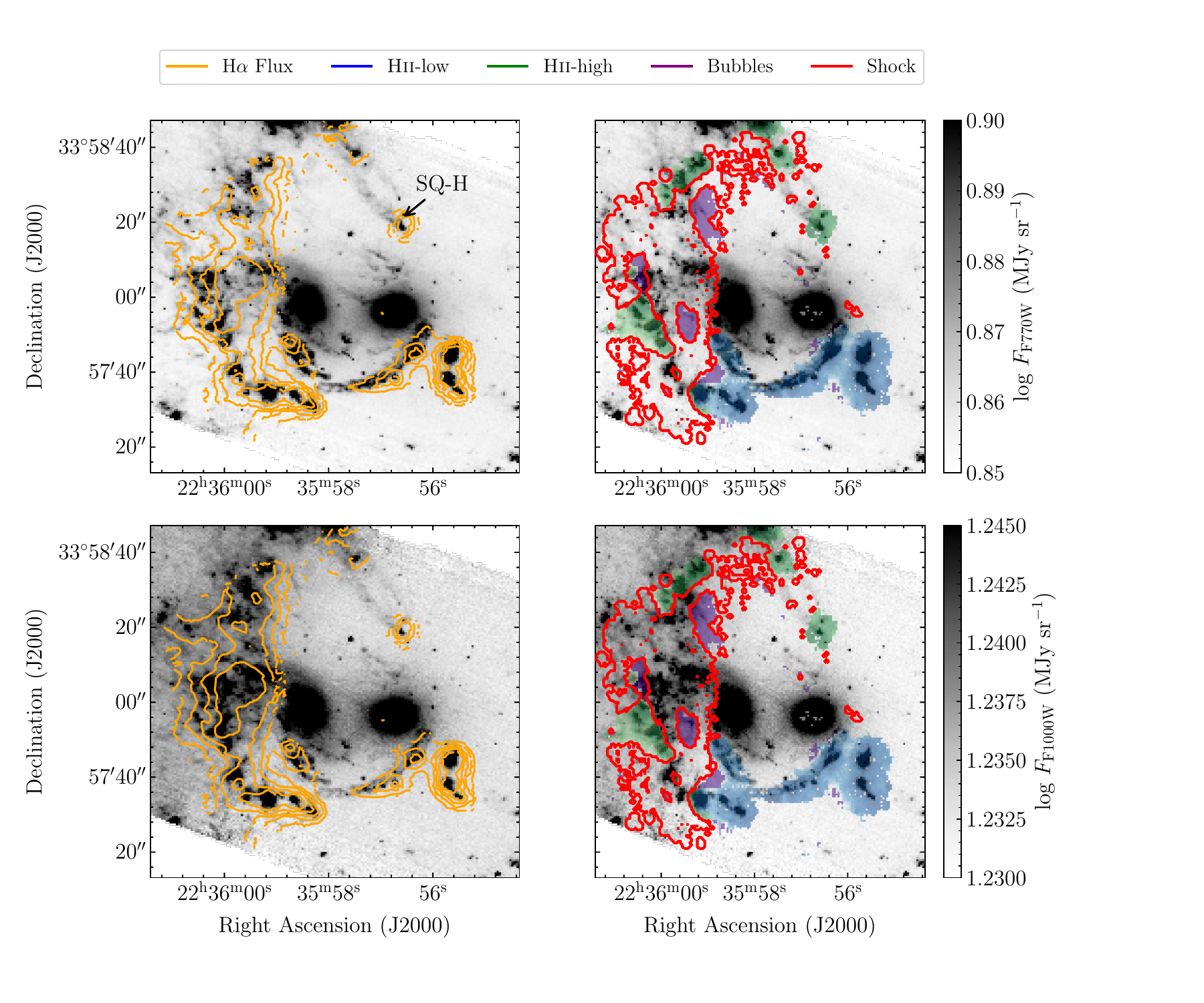}
    \caption{Comparing the H$\alpha$ emission with the PAH and H$_{2}$ emission traced by the \jwst\ MIRI data. The top row shows the F770W \jwst\ image (which traces the PAH emission), while the bottom row shows the F1000W image (which traces the molecular Hydrogen gas) both relative to the flux density scales to the right. The first column shows the MIRI images superimposed with orange contours showing the H$\alpha$ flux distribution, with levels at (1.56, 2.85, 4.29, 6.47 and 10.11)$\,\times\,10^{-16}$\,erg\,s$^{-1}$\,cm$^{-2}$, corresponding to the 10th, 30th, 50th, 70th \&\ 90th percentiles of the H$\alpha$ image. The MIRI images in the second column are overlaid with the kinematically-defined regions from section \ref{sec:weave_overview}, where the \textsc{Hii}-low, \textsc{Hii}-high and bubbles are represented by the blue, green and purple shaded areas, respectively, as in Figure \ref{fig:vel_map}, whereas the shock is outlined in red.}
    \label{fig:Ha_jwst}
\end{figure*}

\begin{figure*}
    \centering
    \includegraphics[trim={2cm 1.5cm 4cm 4cm},clip,width=1.0\textwidth]{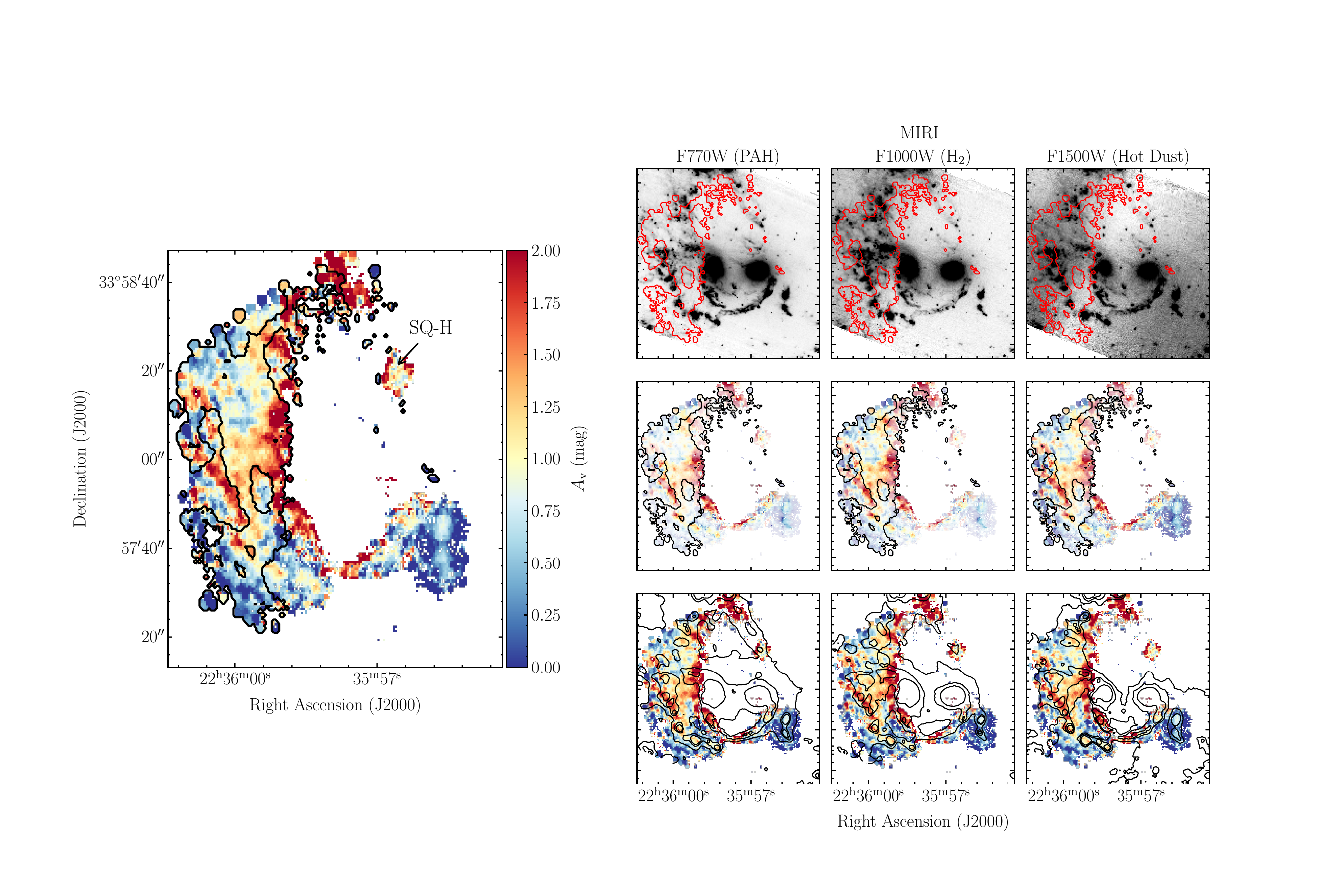}
    \caption{The dust properties of the shocked region surrounding NGC\,7318a \& b. The first panel presents the V-band extinction ($A_{V}$) obtained with the Balmer decrement method, where the black contours indicate the dynamically defined shock region as discussed in Section \ref{sec:weave_overview}. The top row to the right shows the \jwst~MIRI observations centered on 7.7, 10 and 15$\mu$m, where the shocked region is overlaid in red. The middle row represents the $A_{V}$ measurements masked using the intensity of the three images as described in section \ref{sec:dust properties} with the shock overlaid in black.
    The bottom row complements the comparison, where the $A_{V}$ map is superimposed with black contours indicating the 68th, 95th and 99th percentile of each MIRI image. The three columns are therefore used to compare the $V$-band extinction with the PAH complex, H$_{2}$ emission and hot dust continuum, respectively. }
    \label{fig:av_jwst}
\end{figure*}

To compare the properties of the ionised gas with the mid-infrared emission, we make use of the F770W and F1000W \jwst\ MIRI images, which, as mentioned in section \ref{sec:data_jwst}, are effective tracers of PAH and warm H$_{2}$ emission, respectively. To do this, we first convolve the MIRI maps with a Gaussian kernel to match the spatial resolution of the WEAVE data (1.67 arcseconds seeing), and in Figure \ref{fig:Ha_jwst} superimpose orange contours showing the H$\alpha$ emission (in the left column), along with the kinematically-defined regions from section \ref{sec:weave_overview} (in the right column). There is clear correlation between the H$\alpha$ emission and the PAH features visible in the top row, especially in the \textsc{Hii} regions which we identified in the line-of-sight versus velocity dispersion diagram. This is particularly evident for \textsc{Hii}-low (shown in blue in Figure \ref{fig:Ha_jwst}), which coincides with the southern arc, the spiral arm-like structure to the south of NGC7318a\slash b, where previous studies have confirmed that star-formation dominates (e.g. \citealt{iglesias2012new}; \citealt{konstantopoulos2014shocks}; \citealt{puertas2021searching}). However, the correlation is also apparent for \textsc{Hii}-high, which includes the green-shaded regions to the East and North of the shock region (which is outlined in red in the right column of Figure \ref{fig:Ha_jwst}) as well as SQ-H, located to the West. In the bottom row of Figure \ref{fig:Ha_jwst}, on the other hand, there is an apparent association between the H$\alpha$ in the core of the shock region and the H$_2$. As discussed by \citet{guillard2009}, \citet{appleton2017} and \citet{appleton2023multiphase}, this could be evidence of the shock propagating through a multi-phase medium; while the low density regions are heated to X-ray temperatures, higher-density regions can collapse to form H$_2$ on any remaining dust grains that have survived the shock, and this can take place on roughly the dynamical timescale of the collision.
We also note that the dynamically-defined ``bubbles'' appear anti-correlated with both the PAH and H$_{2}$ emission, demonstrating once more behaviour that is distinct from that of both the \textsc{Hii} and shock regions.

To investigate the dust properties of SQ, we use the Balmer decrement method from \cite{dominguez2013dust} to calculate the extinction in the $V$-band ($A_{V}$) across the system for all spaxels with $\ge 3\sigma$ detections in both H$\alpha$ and H$\beta$ emission lines. To do this, we make the standard assumption of case B recombination for a typical gas of temperature of $T = 10^{4}$\,K and an electron number density of $n_{e} = 10^{2}$\,cm$^{-3}$, along with the reddening curve provided by \cite{calzetti2000dust} and an extinction-to-reddening ratio of $R_{V}=4.05$. The resulting $A_{V}$ distribution is shown in the left-hand panel of Figure \ref{fig:av_jwst}, where we can see $V$-band extinction values in the range $0.0 < A_V/\mathrm{mag} < 2.0$. This range is similar to the values obtained for a selection of points along the shock region in SQ by \citet{konstantopoulos2014shocks} and \citet{puertas2021searching}, though we find greater extinction in the shock region than was inferred in the immediate environment of NGC\,7318a \& b by \citet[]{rodriguez2014study}. 

To study this in more detail, in Figure \ref{fig:av_jwst} we put the Balmer-line extinction map in context of the \jwst\ MIRI images, now also including the F1500W image, which traces the hot dust continuum emission (\citealt{appleton2023multiphase}). In the middle row of Figure \ref{fig:av_jwst} we show the $A_V$ map masked using the MIRI images. This was done by defining a colour scale using the $A_{V}$ map, and introducing transparency in inverse proportion to the (base 10) logarithm of the mid-IR flux density in each image (convolved to the WEAVE seeing as in Figure \ref{fig:Ha_jwst}). To complement this comparison, we have further overlaid the $A_{V}$ map with contours representing the 50th, 86th and 95th percentiles of the mid-infrared flux from the three MIRI images. Weighting the $A_V$ map in this manner with the outline of the shock region definition overlaid enables better comparison between the datasets, since it permits us to compare the apparent locations of the obscuring material (as measured by $A_V$) with the distribution of the hot dust and the warm H$_2$ gas \citep[which is thought to form primarily on dust grains; e.g.][]{cazaux2004}.

Several effects are immediately obvious. Firstly, regions with the highest dust obscuration are clearly apparent in each of the top-right panels of Figure \ref{fig:av_jwst}, perhaps associated with the starburst region, SQ-A and along the innermost regions of the structure (i.e. those that are closest in projection to NGC\,7318a \& b), highlighting the presence of gas, dust and PAHs associated with the highest extinction areas, as we might expect. However, the shock itself appears anti-correlated with the highest extinction regions (and this is not a signal-to-noise-ratio effect, since we show only those spaxels with $\ge 3\sigma$ in both Balmer lines), as well as with the \jwst\ dust images. This may suggest that the shock has cleared the region of complex molecules and dust grains, either by sweeping them up as the shock propagates through the medium, or heating them to the point of sublimation \citep[thought to happen in the region of $\sim$1500\,K for carbonaceous grains; e.g.][]{granato1994}. This is an appealing idea, since fast shocks are thought to be the primary means of dust destruction in galaxies \citep[those resulting from supernovae; e.g.][]{jones1994,zhu2019}, although previous works \citep[e.g.][who studied SQ]{guillard2009} have noted that if the shock is weak enough, the dust can survive (and H$_2$ can be formed on it). However, given the complex nature of the system, another possibility is that the ionised and molecular gas have a different spatial distribution on small scales. In this scenario, the shock may be powerful enough to destroy the dust grains in the low-density environment, but in denser, clumpier regions, where the shock velocity is lower, dust can survive, cool and as previously discussed ultimately form H$_{2}$ (see \citealt{guillard2009} for further details). Future high-resolution spectroscopy with \jwst\ will help resolve this.

Other interesting features are apparent: SQ-H (an H\textsc{ii} region) is significantly less visible in the resolution-matched \textit{JWST} F1000W and F15000W maps than in the extinction map. This suggests that SQ-H is either lacking in molecular gas and dust or contains dust sufficiently cool that it is not detected at 15\,$\mu$m. SQ-H is also not detected in the 70\,$\mu$m \textit{Spitzer} data from \citet{xu2008}, implying that the region contains $< 4\times 10^4$\,M$_\odot$ of dust \citep[$5\sigma$ limit obtained assuming an isothermal dust model with $T=30$K and emissivity index $\beta=1.82$ typical for galaxies;][]{smith2013}.

\subsection{The radio morphology of the system} \label{section:radio_morphology}

\begin{figure*}
\centering
\includegraphics[width=2\columnwidth]{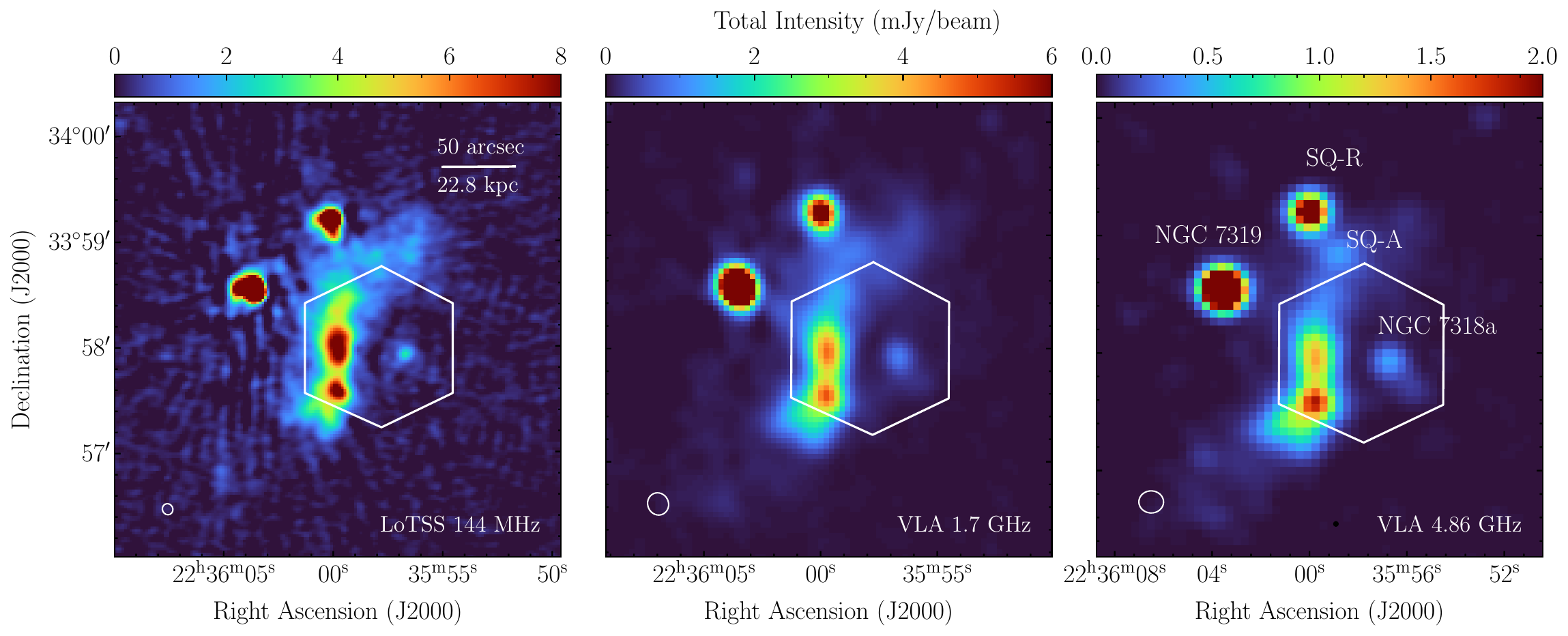}
    \caption{Radio observations of Stephan's Quintet taken by LoTSS at 144\,MHz (left panel) and VLA at 1.7 and 4.86\,GHz (centre and right panels, taken in C- and D- array, respectively). In each panel, the white hexagon indicates the field covered by the WEAVE first-light LIFU observations. The flux scale for each image is indicated by the colourbar at the top of each panel, while the beam size is indicated by the ellipse visible in the lower left corner of each panel. Further details are available in Table \ref{tab:rad_images}.}
    \label{fig:rad_images}
\end{figure*}

The radio images obtained from LoTSS, VLA C-Array 1.7\,GHz, and VLA D-Array 4.86\,GHz are presented in Figure \ref{fig:rad_images}. We can see radio structures consistent with the findings of previous works \citep[e.g.,][]{vanderhulst1981VLA,Aoki1999,Xu2003,Xanthopoulos2004,nikiel2013intergalactic}, but in greater detail as a result of the increased resolution and sensitivity (including to extended emission) of LoTSS relative to previously published data. The diffuse filament of radio continuum associated with the LSSR is prominent in all three figures and exhibits a `boomerang'-like shape. 
The upper region extends for at least 25\,kpc to the North-West, is fainter and more diffuse than the lower region which is associated with the LSSR, and spans approximately 60\,kpc oriented North-South, beyond the FoV of WEAVE (as indicated by the hexagon). Further to the South, we observe a rapid decrease in radio brightness and an increase in lateral spread, eventually terminating around the foreground source NGC\,7320. 

While NGC\,7318b, the new intruder, is not detected in any of our assembled radio frequency data, we identify
three other relatively compact structures in its vicinity. The first of these sources is NGC\,7319, which is the brightest source visible in all three images, and which \cite{Xanthopoulos2004} have suggested is a Seyfert galaxy. The very high resolution VLA and MERLIN images (0.15 arcseconds) from \cite{Xanthopoulos2004} have shown an FRII-like structure for this source, potentially hinting at an interaction of radio jets with the shocked plasma. However, although the core of NGC\,7319 is clearly extended at 144\,MHz, we cannot robustly identify such an FRII-like structure in the data we have assembled. The second of these sources is the radio source SQ-R, which appears extended in the LoTSS images, and is situated to the North of the diffuse radio filament (as indicated in third panel of Figure \ref{fig:rad_images}).
SQ-R is spatially offset from SQ-A by $\sim26$ arcsec ($\sim12$\,kpc) and may be associated with a marginally extended $m_\mathrm{f444w}^\mathrm{AB} \approx 21.1$ source located $\sim 3.8$\,arcsec to the South; coupled with the extension visible at 144\,MHz, we suggest that this source is likely to be a background radio galaxy, unrelated to SQ. While we are unable to discern the presence of SQ-A itself at 144\,MHz since it is embedded within the roughly uniform surface brightness of the extended emission that surrounds it (in projection, at least), it is clearly identifiable as a separate brightness peak in the 4.86\,GHz data. This indicates that SQ-A has a spectral index that differs from that of the extended emission. 
The third source is the galaxy NGC\,7318a. Despite being significantly detected in all three images, the radio emission from NGC\,7318a is faint in comparison to the other compact sources in the region. 

\section{Shock properties}\label{sec:shock properties}

In this section, we now focus on the properties of the shock region, as defined in the velocity vs. velocity dispersion diagram described in Section \ref{sec:weave_overview}.

\begin{figure*}
\centering
\includegraphics[width=1\textwidth]{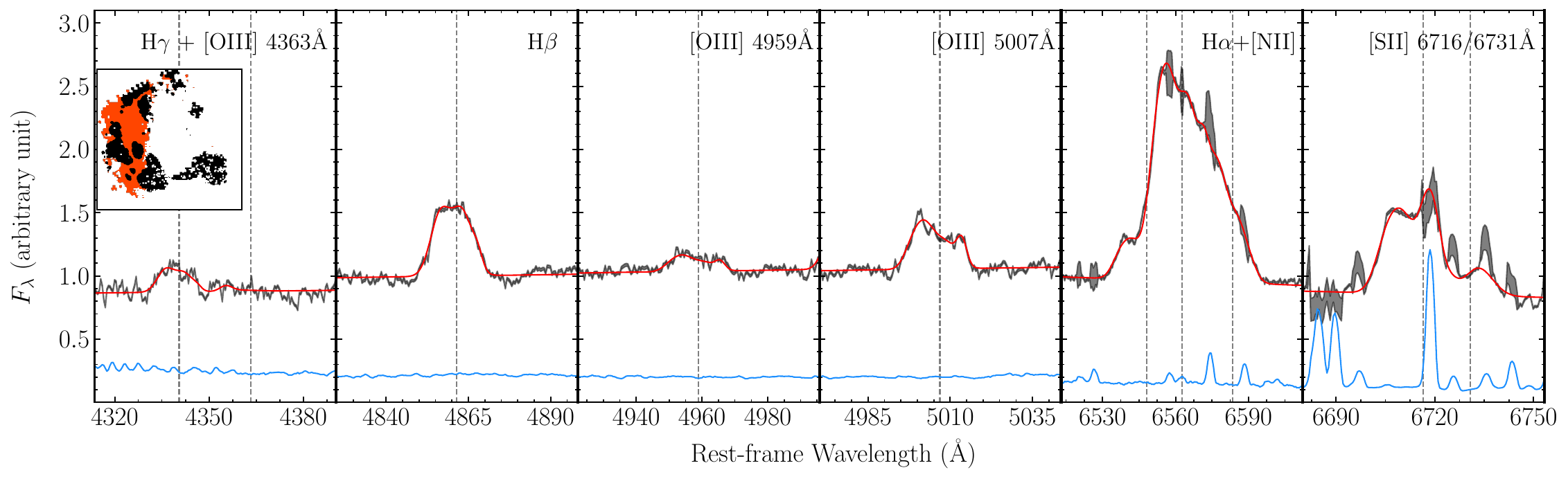}
    \caption{Key wavelength ranges of the stacked spectrum of the shock region (as defined in Section \ref{sec:weave_overview}, shown as red in the embedded plot in the first panel). The different panels show the emission line complexes of interest as indicated in the upper right corner, where the data are represented in grey lines with shaded region enclosing the $1\sigma$ uncertainties, and the best-fit model overlaid in red. The sky model used during the data reduction is shown at the bottom of each panel in blue, demonstrating severe contamination of sky lines at the location of the [S\,\textsc{ii}]$\lambda\lambda$\,6716/6731$\angstrom$ doublet. The black dashed lines denote the wavelength for each emission line in a given complex. The fluxes measured in each line species are detailed in Table \ref{tab:shock_fluxes}.}

    \label{fig:shock_spec}
\end{figure*}

\subsection{The average temperature and number density of the shocked gas} \label{sec:thermodynamics}

To study the shock properties of the system, we aim to use the [\textsc{Oiii}]$\lambda\lambda$ 
(4959+5007)/$\lambda$4363$\angstrom$ line set
and [\textsc{Sii}]$\lambda\lambda$\,6716/6731$\angstrom$ doublet as electron temperature ($T_{e}$) and number density diagnostics ($n_{e}$), respectively (\citealt{osterbrock2006astrophysics}).
Since the [\textsc{Oiii}]$\lambda$4363$\angstrom$ line is not detected in any individual spaxel, we are unable to study variation in the temperature across the shock; however, to determine an average value we use our dynamical definition of the shock region (discussed in Section 4.1) to create a representative stacked spectrum of this region with high signal-to-noise (e.g., \citealt{zhu2015}; \citealt{rigby2018}; \citealt{arnaudova2024}). This was done by summing all spaxels within the dynamically defined shock region, and obtaining a robust measure of the uncertainty on the stack by bootstrapping the summation by sampling the spaxels with replacement (this has the additional benefit of accounting for the correlated noise between neighbouring spaxels; see section \ref{sec:data_weave} for details). The result is shown in Figure \ref{fig:shock_spec}, where each of the six panels shows regions around the locations of emission lines of interest, with the stacked spectrum shown as the grey line with the region representing the derived $1\sigma$ uncertainties shaded. In the bottom of the figure we have also included the (arbitrarily normalised) sky spectrum, indicating the regions suffering from the worst sky line contamination -- particularly around [S\,\textsc{ii}] 6716/6731$\angstrom$. This contamination was not reflected in the inverse variance extension of the reduced data; to reduce the impact of the imperfect sky subtraction on our fits, we have arbitrarily increased the uncertainties on our stacked spectrum by a factor of 10 where prominent sky lines are detected prior to the fitting process. The best-fit model obtained from the MCMC fitting (using a method similar to that described in Section \ref{sec:weave_overview}, however also modelling additional species; [\textsc{Oiii}]$\lambda$4363\AA, the [\textsc{Sii}] doublet, and H$\gamma$ due to its proximity to [\textsc{Oiii}]$\lambda$4363\AA) is overlaid in red. The complicated broad velocity structure is apparent in each emission line complex, and the fluxes measured in each line -- integrated over the four individual Gaussian components for each species --  are detailed in Table \ref{tab:shock_fluxes}. 

Despite stacking the 4,897 spaxels in the shock region, we are unable to detect significant [\textsc{Oiii}]$\lambda4363$\AA\,flux. However, we can still obtain an upper limit on the flux in the [\textsc{Oiii}]$\lambda$4363$\angstrom$ line and thus on the electron temperature. To do this, we use the flux estimates for the three [\textsc{Oiii}] emission lines from the converged MCMC chains and propagate them through the $T_{e}$ equation from \cite{proxauf2014upgrading}, which is an upgraded diagnostic that accounts for Bowen fluorescence and is based on newer atomic data than the standard equation from \cite{osterbrock2006astrophysics}, obtaining a value of $T_{e}<22,500$\,K (at 95 per cent confidence).\footnote{We note that when using the similar relation from \cite{osterbrock2006astrophysics}, the derived temperature is at the upper boundary of where the diagnostic holds ($T_{e}\sim25,000$\,K).}  

At the same time, we use the equation from \citet{proxauf2014upgrading} to estimate the electron density from the [\textsc{Sii}]$\lambda\lambda$\,6716/6731$\angstrom$\ doublet assuming a canonical temperature of $T_{e}=10,000$\,K (consistent with our limit), obtaining $n_{e} = 480\pm70$\,cm$^{-3}$ (or $<$ 695\,cm$^{-3}$ on the upper bound of our temperature range).

Given the range of temperatures permitted based on our limit above, the product of the electron temperature and number density estimate far exceeds the corresponding value for the surrounding plasma traced by the X-rays ($T=0.61\pm0.02$\,keV -- corresponding to $\sim7.1\pm0.2\times 10^{6}$\,K -- and gas density $n=1.167\times10^{-2}$\,cm$^{-3}$ from \citealt{o2009chandra}). Therefore, we cannot determine with confidence the equilibrium state of the system (i.e. whether the shocked gas is under- or over-pressured relative to the surrounding medium), however it is likely that the shock is over-pressure and will expand outwards as the energy dissipates.

\begin{table}
    \caption{Emission line flux measurements obtained for the stacked spectrum of the dynamically defined shock region (see section \ref{sec:weave_overview} for details). The lines are listed in the order of rest-frame wavelength, with fluxes and uncertainties derived from the MCMC chains, where for [O{\sc iii}] $\lambda$4363 we quote an upper limit (at 95 per cent confidence).}
    \label{tab:shock_fluxes}
    \begin{center}

    \begin{tabular}{l|c}
        Line & Flux / $10^{-13}$\,erg cm$^{-2}$ s$^{-1}$ \\
        \hline
        H$\gamma$ & $2.22 \pm  0.06$\\
        
        [O{\sc iii}] $\lambda$4363 & $< 0.38$\\
        
        H$\beta$ & $7.46 \pm  0.16$\\
        
        [\textsc{Oiii}] $\lambda 4959$ & $1.74 \pm  0.03$\\
        
        [\textsc{Oiii}] $\lambda 5007$ & $5.19\pm  0.09$\\
        
        [\textsc{Nii}] $\lambda 6548$ & $4.54 \pm  0.09$\\
        
        H$\alpha$ &  $28.12 \pm  0.55$\\
        
        [\textsc{Nii}] $\lambda 6583$ & $13.6 \pm  0.27$\\
        
        [\textsc{Sii}] $\lambda 6716$ & $7.03 \pm  0.12$\\
        
        [\textsc{Sii}] $\lambda 6731$ & $ 7.21 \pm  0.12$\\
    \end{tabular}
    \end{center}    
\end{table}

\subsection{Emission line ratios}

\subsubsection{Comparing WEAVE data with theoretical shock models}\label{sec:bpt}

To infer additional properties of the shock, we use the line ratios [{O\,\sc iii}] $\lambda$5007$\angstrom$/H$\beta$ and [{N\,\sc ii}] $\lambda$6583$\angstrom$/H$\alpha$ (i.e. the BPT-[\textsc{Nii}] diagram; \citealt{baldwin1981}) using the total line fluxes for all spaxels within the shock region defined in Section \ref{sec:weave_overview}). 
This diagnostic diagram is typically used in the literature to classify sources based on their primary excitation mechanism, often associated with photoionization. In this classification scheme, sources are categorized as star-forming (SF) if they fall below the demarcation line provided by \cite{kauffmann2003}, and as AGN if they lie above the `extreme starburst line' defined in \cite{kewley2006host}. The region in between these lines is referred to as composite, indicating the presence of both mechanisms. However, in the context of this study, where we are dealing with shock-excited gas, we are employing the BPT diagram to compare our observations with theoretical models, and thus obtain additional properties of the shock, such as the metallicity.

To do this, we consider the fast shock models without a precursor from the \textsc{\texttt{MAPPINGS\,III}} library \citep{allen2008} for a pre-shock density of $n=1$\,cm$^{-2}$ (this is the only value available for the models with varying metal abundances, however our conclusions are unchanged if we consider the full range of densities in the libraries),  the full range of magnetic fields (0.0001, 0.5, 1, 2, 3.23, 4, 5 and 10\,$\mu$\,G), and shock velocities between 150\,km\,s$^{-1}$ and 500\,km\,s$^{-1}$, in intervals of 50\,km\,s$^{-1}$, where the differences are most pronounced, and a far higher shock velocity of 900\,km\,s$^{-1}$ for reference (above this value the MAPPINGS library does not include the full range of magnetic field strengths). We do not include shock models with precursors since the model tracks do not occupy the same parameter space as the observed data.

\begin{figure}%
    \centering
    \includegraphics[width=1\columnwidth]{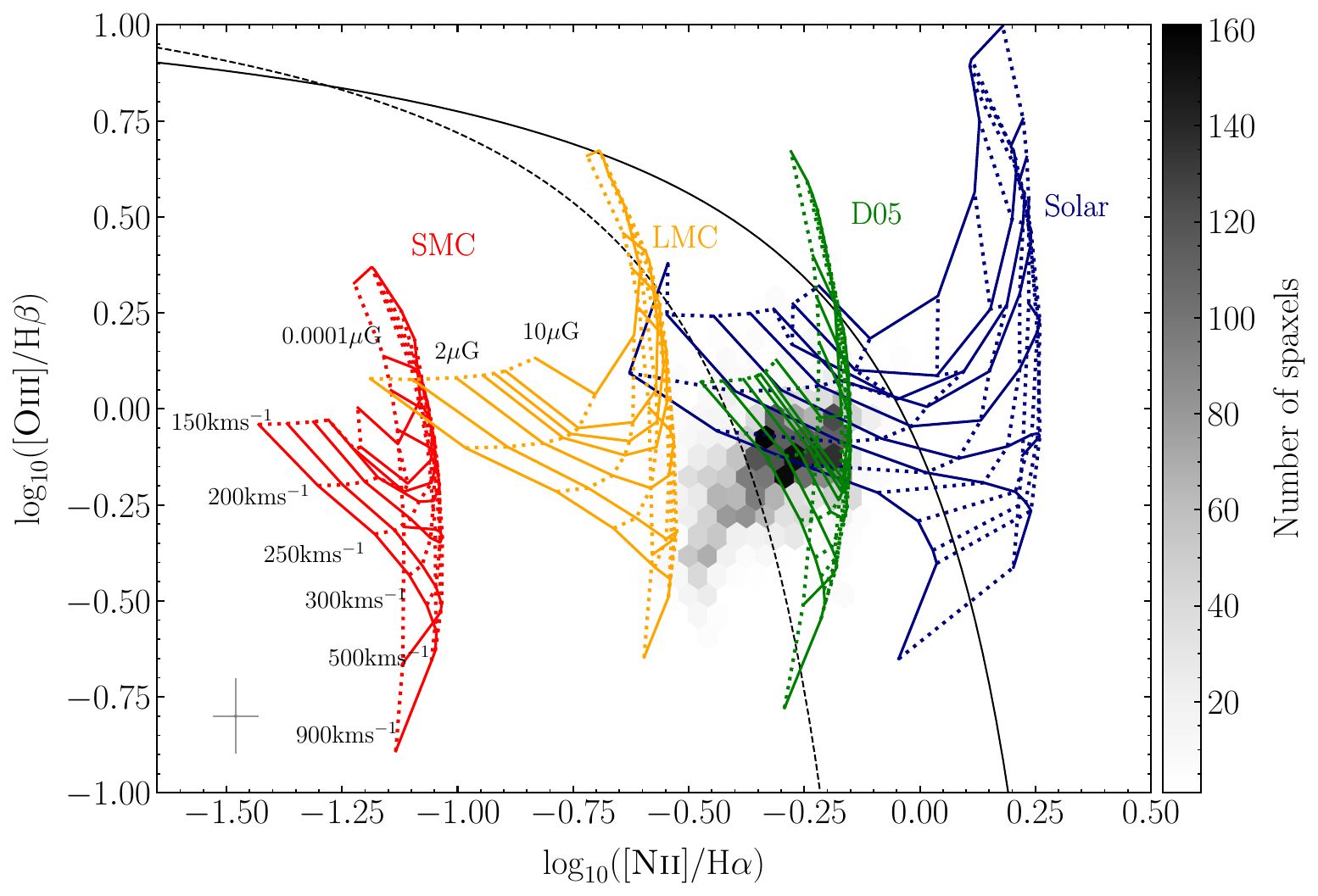}
    \caption{The [{O\,\sc iii}] $\lambda$5007$\angstrom$/H$\beta$ and [{N\,\sc ii}] $\lambda$6583$\angstrom$/H$\alpha$ emission line ratios for all spaxels within the shock region, overlaid with shock models without a precursor of different elemental abundances from the \textsc{\texttt{MAPPINGS III}} library for a pre-shock density of $n=0.1$\,cm$^{-3}$. The shock velocities range from 150 to 500\,kms$^{-1}$ in intervals of 50\,kms$^{-1}$, and a velocity of 900\,kms$^{-1}$, whereas the range of magnetic fields considered is 0.0001, 0.5, 1, 2, 3.23, 4, 5 and 10$\mu$G. The typical error along each axis is represented in the lower left corner. The black solid and dashed lines are the demarcation lines defined by \citet{kewley2001} and \citet{kauffmann2003}. The coloured dotted lines represent constant tracks of shock velocity, whereas the solid lines show constant tracks of magnetic field strengths ($B$). The colourbar denotes the number of spaxels in the shocked region.}
    \label{fig:bpt}
\end{figure}

The results are shown in Figure \ref{fig:bpt}, in which the line ratios in each spaxel are indicated by the 2D histogram in the background (with shading relative to the colourbar at the right). For reference, we have overlaid the two demarcation lines (as solid and dotted lines for the \citealt{kewley2001} and \citealt{kauffmann2003} criteria, respectively). In addition, for the purposes of comparison we have overlaid four shock models, each characterised by different elemental abundances provided in the \textsc{\texttt{MAPPINGS\,III}} library: SMC, LMC, \citeauthor{dopita2005}(\citeyear{dopita2005}; hereafter D05) and Solar. 
It is immediately apparent that the line ratios in the shock are a strong function of the abundances assumed, and that the line ratios in SQ show best agreement with the shock model with D05 abundances, though with some possible overlap with the solar abundance model. 
There appears to be a good agreement between the oxygen and nitrogen abundances in the D05 model (\mbox{[O\slash H]=0.32} and \mbox{[N\slash H]=0.34}) and the metallicity value obtained for the hot X-ray plasma by \citep[$\sim 0.3Z_{\odot}$;][]{o2009chandra}, which could indicate a common origin for the shock and the surrounding medium. However, given that this agreement disappears if different elements are chosen for the comparison (e.g. the D05 value for \mbox{[Fe\slash H]} $\approx 0.01$), this is not a reliable conclusion, especially given that Figure \ref{fig:bpt} also reveals some overlap between the data and the solar abundance \textsc{\texttt{MAPPINGS}} models, which is consistent with the presence of two chemically-different components as implied by previous works (e.g. \citealt{iglesias2012new}; \citealt{puertas2021searching}). 

Taken together, it is clear that better metallicity and abundance diagnostics are required to reliably determine whether the shocked gas in SQ could have a common origin with the plasma in the X-ray emitting halo. Furthermore, when examining the shock velocities, we find that the values obtained depend on the abundance model chosen. While the Solar abundance model suggests that the data are best described by shock velocities within $v_{s}$ = 200-400\,km\,s$^{-1}$ (consistent with previous findings using coarser spatial sampling, but the same method, e.g. \citealt{rodriguez2014study}; \citealt{puertas2021searching}), the overlap with the observations is limited, and the D05 abundance model for example has a similar degree of overlap, but suggests a wider range of velocities. Nevertheless, velocities derived using the Solar abundance models are lower than the difference between the intruder and the systemic group velocity (and we shall return to this point in section \ref{sec:mach number}, below). Given the uncertain abundances, we are unable to make a definitive statement about the shock velocity on the basis of the \textsc{\texttt{MAPPINGS}} models, and instead must use the line-of-sight velocities in the calculations that follow despite the additional source of uncertainty resulting from the possible influence of projection.

\subsubsection{The hardness of the line ratios across the field}
\label{sec:bpt_wider}

\begin{figure*}
    \centering
    \includegraphics[width=.85\textwidth]{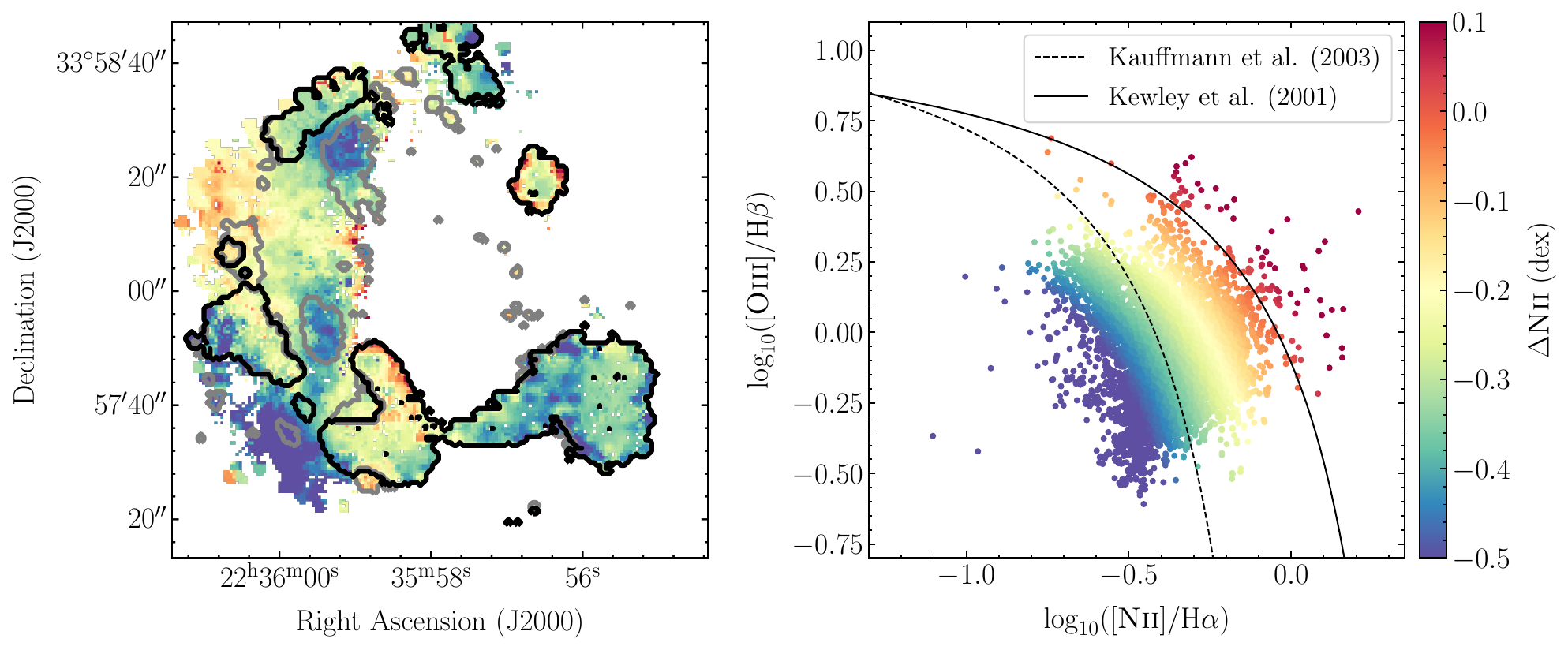}
    \caption{The left panel shows the spatial distribution of all spaxels with a 3$\sigma$ detection in all BPT-[\textsc{Nii}] lines colour-coded based on the orthogonal distance ($\Delta \textsc{Nii}$) to the extreme starburst line from \citet{kewley2001}, as indicated by the right panel. The grey and black contours represent the outline of the bubbles and \textsc{Hii} regions, respectively, with the remaining areas representing the shock region. The right panel shows the BPT-[\textsc{Nii}] diagram, where the solid and dashed lines indicate the demarcation lines from \citet{kewley2001} and \citet{kauffmann2003}, respectively.}
    \label{fig:BPT_spatial}
\end{figure*}

We have also investigated the sky distribution of the full set of spaxels with $> 3\sigma$ detections in all of the BPT lines, according to their orthogonal distance ($\Delta$\textsc{Nii} as in \citealt{dellabruna2022}) to the extreme starburst line from \cite{kewley2001}. Although, as previously discussed, this diagram was originally intended to be used to study galaxies rather than individual spaxels, it can nevertheless provide a widely-used means of investigating the hardness of the observed line ratios in different regions within a galaxy \citep[e.g.][]{iglesias2012new,rodriguez2014study,dellabruna2022}, where spaxels with $\Delta\textsc{Nii}\ga 0$ having harder line ratios (red/orange regions) than those with $\Delta\textsc{Nii}<0$ (purple/green regions), that is closer to and below the \cite{kauffmann2003} line. This can be seen in Figure \ref{fig:BPT_spatial}, where the left panel shows the 2D-BPT distribution, overlaid with contours indicating the kinematically-defined regions from section \ref{sec:weave_overview}, which is colour-coded based on 1D-BPT distribution in the right panel, where the demarcation lines from \cite{kewley2001} and \cite{kauffmann2003} are included. 

The principal result revealed is that each of the kinematically-identified features (`shock', `bubbles' and `\textsc{Hii}-low\slash high') contain spaxels with smoothly varying $\Delta\textsc{Nii}$. The \textsc{Hii}-low\slash high regions generally appear to contain softer line ratios, with $\Delta\textsc{Nii}$ ranging from -0.5 to -0.3. However, the spaxels located in the southern-most end of the shock are associated with the lowest $\Delta\textsc{Nii}$, and these regions are neither \textsc{Hii} regions (on account of the high velocity dispersion), nor associated with the foreground galaxy, NGC 7320 (due the very different line-of-sight velocities; $v_{\mathrm{NGC7320}}$ = 786\,km\,s$^{-1}$; \citealt{yttergren2021gas}). Some of the bubbles also follow this softer trend, but the rest are associated with line ratios produced from harder ionising radiation ($\Delta\textsc{Nii}\sim-0.2$). This is perhaps not a surprise, given the complex nature of the system; we have already shown above that the shock models fill the BPT diagram depending on the elemental abundances, potentially suggesting chemically distinct regions, and other works \citep[e.g.][]{dopita2006, sanchez2015, dellabruna2022} have shown that purely star-forming regions can likewise extend beyond the \cite{kauffmann2003} line and even cross the \cite{kewley2001} line.

\subsection{The plasma in the shock}\label{sec:spix}

The spectra of sources at radio frequencies are often modelled as a power law in frequency, such that $S_\nu \propto \nu^{\alpha}$, with $\alpha$ representing the spectral index. Models of the evolving radio frequency spectra for synchrotron emission \citep[e.g.][]{harwood2017} account for the fact that the spectral index observed between any two frequencies evolves with time. This happens because higher-energy electrons lose energy faster than their lower-energy counterparts, causing radio sources to fade more quickly at higher frequencies. As a result, older radio sources exhibit an increase in spectral curvature (i.e. the difference between the spectral index at high and low frequencies; for a discussion see \citealt{calistro2017}), departing from the spectral index that would have been observed at $t=0$ (the so-called `injection index'). 
Therefore, as low frequencies age more slowly, the low-frequency spectral index can be used as a good proxy for the injection index that otherwise has to be assumed from models. In this context, measuring the spectral curvature observed across a wide range of radio frequency observations can provide useful constraints on the physical conditions of the plasma, for example within the shock in SQ.

To this end, we have generated two spectral index maps by combining the new LoTSS data at 144\,MHz with the VLA data at 1.7\,GHz to calculate $\alpha^{144}_{1.7}$ (our best estimate of the injection index in the SQ shock), and the VLA 1.7 and 4.86\,GHz data to derive the high-frequency spectral index; $\alpha^{1.7}_{4.86}$. To achieve this, we have convolved all three images to a common beam size of $14\times14~\mathrm{arcsec}^2$ and have ensured that they are positionally coincident. The spectral indices as a function of position around SQ were then calculated using the {\tt \textsc{AIPS}} task {\tt \textsc{COMB}}, where pixels with intensity values below 3$\sigma$ were masked to avoid low signal-to-noise regions. 
 
\begin{figure}
    \centering
    \includegraphics[width=0.99\columnwidth]{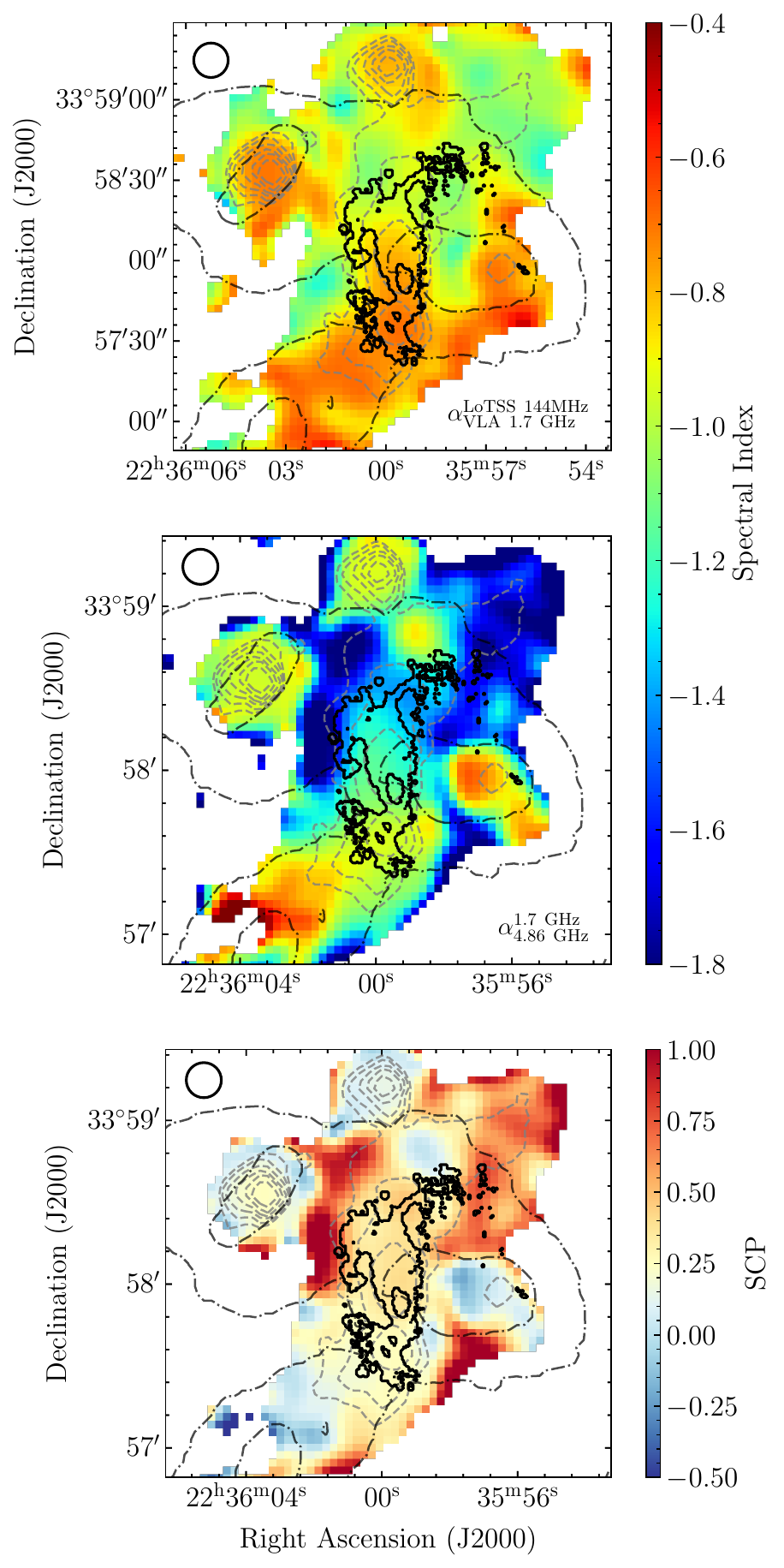}
    \caption{Spectral index image obtained using the LoTSS 144\,MHz and VLA C-Array 1.7\,GHz images (top panel), and VLA C-Array 1.7\,GHz and VLA D-Array 4.86\,GHz images (middle panel), along with the variation of the SCP (bottom panel) across SQ. For context, we have added LoTSS 144\,MHz and \jwst~NIRCam total intensity contours in grey dashed and black dot-dashed curves, respectively. In addition, the outline of the shocked region is included as a solid thick black line. The beam size of the radio frequency images (which have been convolved to be consistent across all frequencies) is indicated by the ellipse in the top left corner.}
    \label{fig:spix_scp}%
\end{figure}

The results for $\alpha_\mathrm{LOW} \equiv \alpha^{144}_{1.7}$ and $\alpha_\mathrm{HIGH} \equiv \alpha^{1.7}_{4.86}$ are shown in the top and middle panels of Figure \ref{fig:spix_scp}, respectively, with the spectral index at each location and between each pair of frequencies indicated by the colourbar to the right. 
The fact that the magnitude of the high-frequency indices is generally larger than the low-frequency indices is immediately apparent, and consistent with expectations for an aged plasma. 
At low frequencies, $\alpha_\mathrm{LOW}$ varies from $-0.73~\pm~0.06$ at the southern extent of the shock to $-0.92\pm0.15$ at the northern extent, with an integrated value of $-0.87\pm0.16$ if we sum flux densities across the entire dynamically defined shock structure. 
On the other hand, the high-frequency indices, $\alpha_\mathrm{HIGH}$, vary from $-0.99\pm0.03$ in the southern end of the shock to $-1.30\pm0.08$ in the northern tip, with an integrated spectral index of $-1.22\pm0.08$ across the entire shock region.  
These spectral index values are consistent with those measured for radio relics (e.g. \citealt{Botteon2020} and references therein), giant synchrotron sources that were generated a significant time ago by shocks crossing their intra-cluster medium (e.g \citealt{feretti2012clusters}).
 
Ageing radio sources tend to exhibit convex spectra \citep[e.g.][]{Chyzy2018,Heesen2022}. The curvature of such convex spectra can be represented using the spectral curvature parameter, defined as $\mathrm{SCP} = \alpha_{\mathrm{LOW}} - \alpha_{\mathrm{HIGH}}$. 
This parameter serves as an indicator of the age of the radio source, where sources with SCP $>0.5$ are typically associated with radio remnants lacking continuous injection of energised plasma (e.g., from an accreting AGN) and are gradually fading away due to radiative and adiabatic losses \citep{Murgia2011,Singh2021}. The variation in SCP across SQ is presented in the bottom panel of Figure \ref{fig:spix_scp}. We see that the SCP across the shock region is low and uniform (median value $\sim 0.32 \pm 0.04$), implying the presence of plasma of similar age throughout the dynamically defined shock region. This region is in fact surrounded by areas with higher SCP values, exceeding 0.5, suggesting that the plasma within the shock is at a different stage of ageing compared to its surroundings, likely the result of the shock energizing or compressing the plasma as discussed further below.
Together, these factors offer support for our decision to define the shock region based on the nebular emission-line fitting.
Several regions with lower (more active) SCP are also apparent, for example the core of NGC\,7318a (the central galaxy to the west of the new intruder), NGC\,7319 (toward the north east of Figure \ref{fig:spix_scp}) as well as SQ-R and SQ-A, suggesting the presence of younger and more energetic plasma with a flatter spectrum, consistent with ongoing particle acceleration due to either star-formation or AGN activity at these locations. 

\subsection{The strength of the shock}\label{sec:mach number}

To determine the strength of the shock propagating through the multi-phase media (the cold gas and the surrounding plasma), we estimate Mach numbers, $\mathcal{M}$, defined as the ratio of the shock velocity to the sound speed in the relevant medium. Both cold (dense) and hot (diffuse) media are of interest; firstly in section \ref{sec:bpt} we showed that the emission line properties of the dynamically-defined `shock' spaxels are consistent with \textsc{\texttt{MAPPINGS}} models in the absence of a photoionised precursor (albeit with a broad possible range of abundances that preclude a definitive velocity estimate) strongly indicating the presence of ionised gas before the shock. Secondly, we are interested in the possible role of the shock in causing the observed radio emission (which we shall investigate in detail in section \ref{sec:adiabatic compression}), and the cosmic rays associated with this emission are almost certainly co-spatial with the volume-filling hot medium rather than with cold neutral gas.

We calculate the sound speed using the standard formula:

\begin{equation}
    c_{s} = \sqrt{\frac{\gamma_{g} k_{\rm{B}} T} {\mu m_{\rm{H}}}},
\end{equation}

\noindent where $\gamma_{g}=5/3$ is the adiabatic index for monatomic gas, $k_{\rm{B}}$ is the Boltzmann constant, $\mu$ represents the mean molecular weight, $m_{\rm{H}}$ is the mass of a hydrogen atom and $T$ corresponds to the temperature of the medium through which the shock is propagating. In this way we obtain sound speeds around 2\,km\,s$^{-1}$ for the cold gas (assuming a typical temperature of 100\,K, e.g. \citealt{sg}), and 440\,km\,s$^{-1}$ for the 0.4\,keV (equivalent to $4.6\times 10^6$\,K) plasma, which is the proposed temperature from \cite{o2009chandra} prior to the collision event (assuming $\mu = 1$ for the neutral gas and 0.5 for the plasma).

We estimate the shock velocity ($v_{s}$) in the cold gas by considering the difference (hereafter $\Delta v$) between the mean velocity from the summed spectrum of the shock region (6268\,km\,s$^{-1}$) and that of the new intruder (5774\,km\,s$^{-1}$), and a pre-shock number density of $n_{\mathrm{H}}$=1\,cm$^{-3}$ (as assumed by the \textsc{\texttt{MAPPINGS III}} models from section \ref{sec:bpt}), following the discussion in \cite{guillard2009}. This results in $v_{s}\sim \sqrt{n_{h}/n_{\mathrm{H}}}\,\Delta v\sim 50$\,km\,s$^{-1}$, where $n_{h}$ is the number density of the hot medium, as measured by \cite{o2009chandra}.
This indicates that the shock propagating through the cold gas is very strong, irrespective of projection effects ($\mathcal{M}_\mathrm{Cold} \ga 25$). 
Such a shock would be more than sufficient to excite the neutral medium into the ionized gas observed with WEAVE, providing clear evidence of its origin. In addition, the hypersonic nature of the shock suggests that -- if any dust can survive -- it is likely confined to only the most dense regions.

A similar calculation for the shock velocity in the hot X-ray emitting gas (based on the difference between the group systemic radial velocity of 6,600\,km\,s$^{-1}$, to which the hot gas is presumably bound, and that of the new intruder) gives $\mathcal{M}_\mathrm{Hot} \ga 1.9$. Previous studies (e.g. \citealt{trinchieri2003}; \citealt{o2009chandra}) have suggested that the shock may be oriented at $\sim31{^\circ}$ to the line of sight, in which case $\mathcal{M}_\mathrm{Hot}\sim3.8$. Irrespective of the geometry, it is clear that the shock propagating through the X-ray plasma is relatively weak ($\mathcal{M}\lesssim3-5$; e.g \citealt{botteon2020shock}). These values (like the magnetic field strengths, discussed in Section \ref{sec:bpt}) are consistent with literature values for internal shocks associated with turbulent flow motions in the ICM \citep[e.g.,][]{Ryu2008,Vazza2012,Brunetti2014}, and are far smaller than those typically associated with accretion onto clusters of galaxies \citep[e.g.][]{skillman2008}. Some works have suggested that shocks with Mach numbers this low are unlikely to generate radio emission from the hot, ionized phase via particle acceleration (e.g. \citealt{vink2014}), and this is consistent with the fact that the radio spectral indices in the shocked region, particularly at high frequencies, are relatively steep, whereas we would expect them to be reset to a flat spectrum by efficient particle acceleration. Instead, the shock is more likely to affect the radio surface brightness by compressing the medium, a possibility we will now consider.

\subsection{Adiabatic compression}\label{sec:adiabatic compression}

In the previous section, we found that the Mach number of the shock ($\mathcal{M}\sim3.8$) in the hot gas is low, suggesting that the collision of NGC\,7318b has generated a weak shock in the hot X-ray medium. Such shocks are inefficient at accelerating the supra-thermal tail of electrons in the hot medium up to relativistic energies \citep{Ensslin2001,Kang2007}. Instead, the principal driver of the increase in the synchrotron emission associated with the shock is believed to be the adiabatic compression of the pre-existing cosmic ray electron population, which is mixed in with the hot phase where it can persist for long periods since radiative losses dominate \citep{Ensslin2001, Ensslin2002, Markevitch2005, Markevitch2007,vink2014}. Works led by \cite{Ensslin2001}, \cite{Markevitch2005}, \cite{Cawthorne2006}, and \cite{Colafrancesco2017} have demonstrated that such an event leads to a boost in the radio emission. To test this scenario, we follow \cite{Colafrancesco2017} and calculate the boosting factor ($A$) at 1.4\,GHz as:
\begin{equation} \label{eq:cfact_boost}
A \sim C^{(-s+2)/3}\times[C^{2/3}]^{1-\alpha}\times C^{-1}.
\end{equation}
Here $C$ is the compression ratio of the gas caused by the shock, $\alpha$ is the radio spectral index (which we take to be $\alpha \equiv \alpha_{\rm{LOW}} = -0.85$) and $s$ is the momentum spectral index, given by $s=2\alpha-1$ (e.g. \citealt{longair2011high}). The compression ratio is related to $\mathcal{M}$ and the adiabatic index ($\gamma_g$) by the following expression (e.g. \citealt{Markevitch2007}, \citealt{vanWeeren2017}):
\begin{equation} \label{eq:mach_cfact}
    \mathcal{M} = \left[\frac{2C}{\gamma_g+1 - C(\gamma_g - 1)}\right]^{\frac{1}{2}}
\end{equation}
By taking $\gamma_g = 5/3$ (monatomic ideal gas) and $\gamma_g = 4/3$ (relativistic gas) as the limiting cases for the plasma, we find that $C$ ranges between 3.3 and 4.9, leading to a boosting factor of $A\sim$ 8.6--17.5. 

In the absence of 1.4\,GHz radio flux measurements, we consider the 1.7\,GHz flux density outside the dynamically defined shock region (Section \ref{sec:weave_overview}) and inside the boundary to represent the flux density before and after the shock-induced compression. Comparing these values, we find that the observed boost in 1.7\,GHz flux density is $\sim$10. The broad agreement between the theoretical and observational boost demonstrates that the increase in synchrotron emission can be explained by considering a scenario in which a weak shock adiabatically compresses the existing radio plasma. Adiabatic compression would preserve the curvature of the pre-existing electron energy spectrum while shifting its energy and number normalization to larger values, consistent with the curved spectrum that we observe.

\subsection{Magnetic field strength and shock lifetime}

Assuming equipartition of energy between the relativistic radiating particles and the magnetic field in which they live, we can establish further constraints on the conditions within the shock region. To do this we use the python package \texttt{\textsc{PySynch}} \citep{Hardcastle1998}\footnote{\url{https://github.com/mhardcastle/pysynch}} for calculating the equipartition parameters, which implements the formulae proposed by \cite{Myers1985} and later revised by \cite{Beck2005}. 
Following \cite{Hardcastle1998}, we assume that the electron energy distribution follows a power law in Lorentz factor ($\gamma$) between $\gamma_\mathrm{min}=1$ and $\gamma_\mathrm{max}=1\times10^{5}$. The exponent $s$ is related to the injection index of the synchrotron radio spectra $\alpha_\mathrm{inj}$ as $s = 2\alpha_\mathrm{inj} - 1$. By considering the lowest measured radio spectral index once more, we obtain $s = -2.7$.
We model the shock region as an isotropic cylinder with a radius of 6\,kpc and a height of 35\,kpc. The volume filling factor of the radio plasma in the shock is assumed to be unity and the kinetic energy density of protons is assumed to be negligible when compared to that of the electrons ($\kappa=0$). We use 144\,MHz as the reference frequency in \texttt{\textsc{PySynch}} as it is the least affected by spectral ageing, and by using the integrated flux density of the shock region ($=136$\,mJy), we calculate the equipartition magnetic field strength of $B_\mathrm{eq}$ = 9.5\,$\mu$G. This estimate is once more consistent with values typically found for radio relics (e.g. \citealt{Ryu2008}; \citealt{feretti2012clusters}), as well as the value obtained in \cite{xu2003physical}, albeit with different assumptions. 
However, our model is more physically motivated since the geometry is associated with the dynamical decomposition of ionised gas, and the injection index is determined from the inclusion of the 144\,MHz observation, which is discussed for the first time.

Additionally, we use the model proposed by \cite{vanderLaan1969} to put an upper limit on the radio emission lifetimes of the boosted relativistic electrons. We assume that the initial interaction of the shock front and the SQ system supplied the relativistic electrons with sufficient energy (the ``generation'' phase) for a subsequent long-term radio emission phase (the ``remnant'' phase). During the remnant phase, the energy supply is switched off and the electrons age, undergoing synchrotron and inverse Compton losses (the latter due to interaction with the cosmic microwave background photons). The duration of this remnant phase at frequency $\nu$ 
can be calculated using Equation \ref{eq:electron-lifetime}:
\begin{equation} \label{eq:electron-lifetime}
    \tau \simeq 2.6\times 10^4 \frac{B_\mathrm{eq}^{1/2}}{B_\mathrm{eq}^2+B_\mathrm{R}^2} \big[ (1+z) \nu \big]^{-1/2}\text{ yr},
\end{equation} 
where $B_\mathrm{R}~=4(1+z)^2~\mu\mathrm{G}$ is the equivalent magnetic field of the cosmic microwave background (CMB). 
The tightest constraints on the electron lifetime are provided by the highest-frequency observations (as those frequencies fade the fastest); since the shock is still clearly visible at 4.86\,GHz, this analysis suggests that the age of the shock is $\tau \approx 11$\,Myr, similar to our estimate of the dynamical crossing time for NGC\,7318b ($\sim 14$\,Myr assuming the same geometry as above, and a relative velocity of $826$\,km\,s$^{-1}$ -- see Section \ref{sec:mach number}). Since these numbers are comparable, we suggest that the collision is the likely cause of the shock.

\section{Conclusions}\label{sec:conclusions}

In this work, we have combined IFU observations from the first-light data of the new WEAVE facility with low-frequency radio observations from the LOFAR Two-metre Sky Survey to study the large-scale shock region (LSSR) in Stephan's Quintet. The combination of the arcminute-scale FoV (90 $\times$ 78 arcsec$^{2}$) and high spectral resolution ($R \sim 2500$) across the wavelength range of $3660-9590$\AA~provided by the large IFU mode of WEAVE has enabled us to perform robust spectral line fitting of the system in greater detail than previously possible. With models including up to four velocity components if required by the data, we have been able to de-blend prominent emission lines and study the detailed kinematics of SQ. These results have enabled us to provide a dynamical definition of the shock region based its location on the line-of-sight velocity versus velocity dispersion diagram and estimate its physical properties by combining the WEAVE first light data cube with new 144\,MHz LoTSS observations and additional auxiliary information from the VLA and \jwst.

Our findings can be summarised as follows:
\begin{enumerate}
    \item The shock region contains ionized gas of density $n_{e} = 480 \pm 70$\,cm$^{-3}$ (with a small temperature dependence) and temperature, $T_{e} < 22,500$\,K, located between the \textsc{Hi} filaments identified in \cite{williams2002vla}. The location of the ionised gas, in combination with the hypersonic Mach number of the shock propagating through the pre-existing \textsc{Hi} gas and consistency with the fast shock models from \citet{allen2008} in the absence of a photoionised precursor, strongly suggest that the shock is responsible for the excitation and ionization of the neutral hydrogen, providing a possible explanation for the \textsc{Hi} deficiency in the system (as previously suggested by e.g. \citealt{iglesias2012new}).
  
     \item The metallicity of the shock region obtained by comparing the [{O\,\sc iii}] $\lambda$5007$\angstrom$/H$\beta$ and [{N\,\sc ii}] $\lambda$6583$\angstrom$/H$\alpha$ emission line ratios with the fast shock models from \cite{allen2008} may be consistent with a common origin with the hot X-ray plasma, however for a more definitive statement, or to be able to determine shock velocities from the emission line ratios, more precise elemental abundances in the shocked gas are required. 

     \item The kinematically-defined shock region shows an anti-correlation with areas of highest dust extinction ($A_{V}$) and with \jwst\ MIRI images which trace the PAH, H$_{2}$, and hot dust emission. Given the hypersonic nature of the shock in the \textsc{Hi} medium, we propose that if any dust has been able to survive the shock, it will be dust grains in the densest regions;  those in lower density environments have almost certainly been destroyed. 
   
    \item The 144\,MHz LOFAR observations combined with 1.7 and 4.86\,GHz VLA observations show that the shock is consistent with containing a homogeneous relativistic particle population with age $\lessapprox $11\,Myr, similar to the crossing time of the new intruder. We conclude that the passage of NGC\,7318b is the likely cause of the 35\,kpc shock. 

    \item The shock propagating through the pre-existing hot X-ray plasma, where the radio emission is generated, is found to be relatively weak. The low Mach number ($\mathcal{M}$ = 3.8) suggests that the shock is unable to efficiently accelerate particles. Instead, we propose that it is adiabatically compressing the medium, by showing that the theoretical boost in radio emission (a factor of $\sim 8.6-17.5$) is compatible with the observed one ($\sim 10$). Combining this information with the spectral curvature parameter (which is highly uniform over the region which we identify as the shock on the basis of the nebular line dynamics), we propose that the shock has compressed a pre-existing radio-emitting plasma with a curved radio spectrum, likely the product of previous interactions.
 
\end{enumerate}

Overall, in addition to the new understanding of SQ, this work highlights the sort of studies that are now becoming possible by combining data from the newly-commissioned WEAVE facility \citep{dalton2012weave} with exciting new and archival multi-wavelength datasets. This approach is central to many of the WEAVE surveys, described by \citet{Jin2023}, which will spend an initial period of five years pursuing a range of exciting science (both Galactic and extragalactic). The ability of WEAVE observations to unlock the huge diagnostic power of the LOFAR observatory forms a key motivation for the WEAVE-LOFAR survey \citep{smith2016}, which will obtain $> 1$\,million optical spectra of 144\,MHz sources identified in LoTSS \citep[e.g.][]{shimwell2022lofar,hardcastle2023}.

\section*{Acknowledgements}
The authors sincerely thank the reviewer (Dr Phil Appleton) for a report which greatly improved the quality of this paper.
MIA acknowledges support from the UK Science and Technology Facilities Council (STFC) studentship under the grant ST/V506709/1.
SD acknowledges support from the STFC studentship via grant ST/W507490/1.
MIA, DJBS and MJH acknowledge support from the STFC under grant ST/V000624/1. DJBS also acknowledges support from STFC under grant ST/Y001028/1.
RJS acknowledges support from the STFC through grants ST/T000244/1 and ST/X001075/1
KJD acknowledges support from the STFC through an Ernest Rutherford Fellowship (grant number ST/W003120/1).
PNB is grateful for support from the UK STFC via grant ST/V000594/1.
SS acknowledges support from the STFC via studentship grant number ST/X508408/1.
RGB acknowledges financial support from the Severo Ochoa grant CEX2021-001131-S funded by MCIN/AEI/ 10.13039/501100011033 and grant PID2022-141755NB-I00.
JHK acknowledges grant PID2022-136505NB-I00 funded by MCIN/
AEI/10.13039/501100011033 and EU, ERDF.
KMH acknowledges financial support from the grant PID2021-123930OB-C21 funded by MICIU/AEI/ 10.13039/501100011033 and by ERDF/EU.
EE acknowledges support from UKRI FLF (MR/T042842/1).
TB acknowledges support from the project grant No. 2018-04857 from the Swedish Research Council.
AM acknowledges financial support through grants PRIN-MIUR 2020SKSTHZ and NextGenerationEU" RFF M4C2 1.1 PRIN 2022 project 2022ZSL4BL INSIGHT.
AI and ML acknowledge financial support from INAF Mainstream grant 2019 WEAVE StePS 1.05.01.86.16 and INAF Large Grant 2022 WEAVE StePS 1.05.12.01.11.
JMA acknowledges the support of the Viera y Clavijo Senior program funded by ACIISI and ULL and the support of the Agencia Estatal de Investigaci\'on del Ministerio de Ciencia e Innovaci\'on (MCIN/AEI/10.13039/501100011033) under grants nos. PID2021-128131NB-I00 and CNS2022-135482 and the European Regional Development Fund (ERDF) ‘A way of making Europe’ and the ‘NextGenerationEU/PRTR’
DA acknowledges financial support from the Spanish Ministry of Science and Innovation (MICINN) under the 2021 Ramón y Cajal program MICINN RYC2021‐032609.
SRB acknowledges financial support by NextGeneration EU/PRTR and MIU (UNI/551/2021) through grant Margarita Salas-ULL, and from the Spanish Ministry of Science and Innovation (MICINN) through the Spanish State Research Agency through grants PID2021-122397NB-C21, and the Severo Ochoa Programme 2020-2023 (CEX2019-000920-S). JALA acknowledges financial support from MCIU through the grant PID2023-153342NB-I00.
This work was (partially) supported by the Spanish MICIN/AEI/10.13039/501100011033 and by "ERDF A way of making Europe" by the “European Union” through grant PID2021-122842OB-C21, and the Institute of Cosmos Sciences University of Barcelona (ICCUB, Unidad de Excelencia ’Mar\'{\i}a de Maeztu’) through grant CEX2019-000918-M.
This work was supported by the Programme National Cosmology et Galaxies (PNCG) of CNRS/INSU with INP and IN2P3, co-funded by CEA and CNES, and ANR under contract ANR-22-CE31-0026. 
The WEAVE core data processing is carried out by the Cambridge Astronomical Survey Unit (CASU) at the Institute of Astronomy, University of Cambridge (supported by UKRI-STFC grants: ST/M003175/1, ST/M007626/, ST/P003486/1, ST/N005805/1, ST/T003081/1 and ST/X001857/1). 
All INAF researchers acknowledge the support from INAF grant 1.05.03.04.05.
Co-funded by the European Union. Views and opinions expressed are however those of the author(s) only and do not necessarily reflect those of the European Union. Neither the European Union nor the granting authority can be held responsible for them. 

Funding for the WEAVE facility has been provided by UKRI STFC, the University of Oxford, NOVA, NWO, Instituto de Astrofísica de Canarias (IAC), the Isaac Newton Group partners (STFC, NWO, and Spain, led by the IAC), INAF, CNRS-INSU, the Observatoire de Paris, Région Île-de-France, CONCYT through INAOE, the Ministry of Education, Science and Sports of the Republic of Lithuania, Konkoly Observatory (CSFK), Max-Planck-Institut für Astronomie (MPIA Heidelberg), Lund University, the Leibniz Institute for Astrophysics Potsdam (AIP), the Swedish Research Council, the European Commission, and the University of Pennsylvania.  The WEAVE Survey Consortium consists of the ING, its three partners, represented by UKRI STFC, NWO, and the IAC, NOVA, INAF, GEPI, INAOE, Vilnius University, FTMC – Center for Physical Sciences and Technology (Vilnius), and individual WEAVE Participants. Please see the relevant footnotes for the WEAVE website\footnote{\href{https://weave-project.atlassian.net/wiki/display/WEAVE}{https://weave-project.atlassian.net/wiki/display/WEAVE}} and the full list of granting agencies and grants supporting WEAVE\footnote{\href{https://weave-project.atlassian.net/wiki/display/WEAVE/WEAVE+Acknowledgements}{https://weave-project.atlassian.net/wiki/display/WEAVE/\\WEAVE+Acknowledgements}}.

LOFAR is the Low Frequency Array designed and constructed by ASTRON. It has observing, data processing, and data storage facilities in several countries, which are owned by various parties (each with their own funding sources), and that are collectively operated by the ILT foundation under a joint scientific policy. The ILT resources have benefited from the following recent major funding sources: CNRS-INSU, Observatoire de Paris and Université d’Orléans, France; BMBF, MIWFNRW, MPG, Germany; Science Foundation Ireland (SFI), Department of Business, Enterprise and Innovation (DBEI), Ireland; NWO, The Netherlands; The Science and Technology Facilities Council, UK; Ministry of Science and Higher Education, Poland; The Istituto Nazionale di Astrofisica (INAF), Italy.

This work is based, in part, on observations made with the NASA/ESA/CSA James Webb Space Telescope. The data were obtained from the Mikulski Archive for Space Telescopes at the Space Telescope Science Institute, which is operated by the Association of Universities for Research in Astronomy, Inc., under NASA contract NAS 5-03127 for JWST. These observations are associated with program $\#2732$.

This research made use of the Dutch national e-infrastructure with support of the SURF Cooperative (e-infra 180169) and the LOFAR e-infra group. The Jülich LOFAR Long Term Archive and the German LOFAR network are both coordinated and operated by the Jülich Supercomputing Centre (JSC), and computing resources on the supercomputer JUWELS at JSC were provided by the Gauss Centre for Supercomputing e.V. (grant CHTB00) through the John von Neumann Institute for Computing (NIC). This research made use of the University of Hertfordshire high performance computing facility and the LOFAR-UK computing facility located at the University of Hertfordshire and supported by STFC [ST/V002414/1], and of the Italian LOFAR IT computing infrastructure supported and operated by INAF, and by the Physics Department of Turin University (under an agreement with Consorzio Interuniversitario per la Fisica Spaziale) at the C3S Supercomputing Centre, Italy.

The authors thank Lourdes Verdes-Montenegro for providing the published HI data.
The authors acknowledge the Spanish Prototype of an SRC (SPSRC) service and support funded by the Ministerio de Ciencia, Innovación y Universidades (MICIU), by the Junta de Andalucía, by the European Regional Development Funds (ERDF) and by the European Union NextGenerationEU/PRTR. The SPSRC acknowledges financial support from the Agencia Estatal de Investigación (AEI) through the "Center of Excellence Severo Ochoa" award to the Instituto de Astrofísica de Andalucía (IAA-CSIC) (SEV-2017-0709) and from the grant CEX2021-001131-S funded by MICIU/AEI/ 10.13039/501100011033.

M. I. Arnaudova led the analysis and the writing of the manuscript;
S. Das, D. J. B. Smith, M. J. Hardcastle, N. Hatch and S. C. Trager made defining contributions to the analysis and to the writing of the manuscript;
M. Balcells, G. B. Dalton, S. C. Trager, L. Domínguez-Palmero and C. Fariña designed and executed the observations;
R. J. Smith, D. Aguado, J. C. McGarry, S. Shenoy, J. P. Stott, J. H. Knapen, K. M. Hess, K. J. Duncan, A. Gloudemans, P. N. Best, R. Garcia-Benito, R. Kondapally, M. Balcells, G. S. Couto, E. L. Escott and S. Jin contributed to the writing of the manuscript;
J. M van der Hulst performed the initial explorations of the first-light data cubes;
D. N. A. Murphy, A. Molaeinezhad, M. J. Irwin, C. C. Worley, S. Hughes, G. D’Ago, N. A. Walton and E. Molinari managed, developed and executed the data processing and reduction pipelines;
S. Das and J. C. McGarry reprocessed archival VLA data;
K. M. Hess reprocessed the HI data from Williams et al. (2002);
G. B. Dalton, S. C. Trager, D. C. Abrams, J. A. L. Aguerri, A. Vallenari, C. R. Benn and S. Jin envisioned and managed the realisation of the overall design concept of the WEAVE facility;
D. Cano-Infantes, P. J. Concepción, G. B. Dalton, K. Dee, E. Gafton, F. J. Gribbin, S. Hughes, I. J. Lewis,  S. Picó,  A. W. Ridings, E. Schallig, J. Skvarč and R. Stuik are members of the WEAVE Instrument Team and made key contributions towards the building of the WEAVE instrument;
D. Aguado, R. Barrena, T. Bensby, P. N. Best, D. Bettoni, R. Carrera, G. S. Couto, A. B. Drake, J. E. Drew, K. J. Duncan, M. Fossatti, M. Fumagalli, R. García-Benito, A. Gloudemans, M. J. Hardcastle, N. Hatch, K. M. Hess, A. Iovino, S. Jin, R. Kondapally, M. Longhetti, J. Méndez-Abreu, A. Mercurio, M. Mongui\'{o}, M. Pieri, S. Rodriguez-Berlanas, M. Romero-G\'{o}mez, T. Shimwell, D. J. B. Smith, J. P. Stott, S. C. Trager and A. Vallenari made key contributions towards the planning of the WEAVE surveys that form the backbone of the WEAVE science case, thereby also informing the design of the WEAVE instrument and overall facility.

\section*{Data Availability}
The fully reduced WEAVE data cubes for this study will be available from the WEAVE Archive System (\url{http://portal.was.tng.iac.es/}).
The radio data used in this work are publicly available and can be found at the LOFAR Surveys (\url{www.lofar-surveys.org}), and at the NRAO Archive Interface (\url{https://data.nrao.edu/portal/}).



\bibliographystyle{mnras}
\bibliography{ref} 


\newpage
\appendix

\section{Author Affiliations} \label{sec:affiliations}
$^{1}$\footnotesize{\textit{Centre for Astrophysics Research, University of Hertfordshire, College Lane, Hatfield AL10 9AB, UK}} \\
$^{2}$\footnotesize{\textit{School of Physics and Astronomy, University of Nottingham, University Park, Nottingham NG7 2RD, UK}} \\
$^{3}$\footnotesize{\textit{Kapteyn Astronomical Institute, University of Groningen, Postbus 800, 9700 AV Groningen, The Netherlands}} \\
$^{4}$\footnotesize{\textit{Centre for Extragalactic Astronomy, Department of Physics, Durham University, Durham DH1 3LE, UK}} \\
$^{5}$\footnotesize{\textit{Department of Physics, Lancaster University, Lancaster LA1 4YB, UK}} \\
$^{6}$\footnotesize{\textit{Instituto de Astrof\'{i}sica de Canarias, Calle V\'{i}a L\'{a}ctea s/n,  38205 La Laguna, Tenerife, Spain}} \\
$^{7}$\footnotesize{\textit{Departamento de Astrof\'{i}sica, Universidad de La Laguna, 38206 La Laguna,
Tenerife, Spain}} \\
$^{8}$\footnotesize{\textit{Department of Space, Earth and Environment, Chalmers University of Technology, Onsala Space Observatory, 43992 Onsala, Sweden}} \\
$^{9}$\footnotesize{\textit{ASTRON, Netherlands Institute for Radio Astronomy, Oude Hoogeveensedijk 4, 7991 PD, Dwingeloo, The Netherlands}} \\
$^{10}$\footnotesize{\textit{Instituto de Astrof\'{i}sica de Andaluc\'{i}a, CSIC, Glorieta de la Astronom\'{i}a s/n,
18008 Granada, Spain}} \\
$^{11}$\footnotesize{\textit{Institute for Astronomy, University of Edinburgh, Royal Observatory, Blackford Hill, Edinburgh, EH9 3HJ, UK}} \\
$^{12}$\footnotesize{\textit{NSF's NOIRLab, Gemini Observatory, 670 N A\'{o}hoku Place, HI-96720, Hilo, USA}} \\
$^{13}$\footnotesize{\textit{Isaac Newton Group of Telescopes, Apartado 321, 38700 Santa Cruz de la Palma, Spain}} \\
$^{14}$\footnotesize{\textit{Leibniz-Institut f\"{u}r Astrophysik Potsdam, An der Sternwarte 16, 14482, 
Potsdam, Germany}} \\
$^{15}$\footnotesize{\textit{Lund Observatory, Division of Astrophysics, Department of Physics, Lund University, Box 118, SE-22100 Lund, Sweden}} \\
$^{16}$\footnotesize{\textit{INAF, Osservatorio Astronomico di Padova, Vicolo Osservatorio 5, 35122 Padova, Italy}} \\
$^{17}$\footnotesize{\textit{INAF, Osservatorio di Astrofisica e Scienza dello Spazio di Bologna, via P. Gobetti 93/3, 40129, Bologna, Italy}} \\
$^{18}$\footnotesize{\textit{Department of Physics, University of Oxford, Keble Rd, Oxford OX1 3RH, UK}} \\
$^{19}$\footnotesize{\textit{Institute of Astronomy, University of Cambridge, Madingley Road, Cambridge CB3 0HA, UK}} \\
$^{20}$\footnotesize{\textit{Department of Physics \& Astronomy, University College London, Gower Street, London WC1E 6BT, UK}} \\
$^{21}$\footnotesize{\textit{Universit\`{a} degli studi di Milano-Bicocca, Piazza della scienza 3, I-20126 Milano, Italy}} \\
$^{22}$\footnotesize{\textit{INAF, Osservatorio Astronomico di Brera, via Brera 28, I-20121 Milano, Italy}} \\
$^{23}$\footnotesize{\textit{Dipartimento di Fisica G. Occhialini, Universit\`{a} degli Studi di Milano Bicocca, Piazza della Scienza 3, I-20126 Milano, Italy}} \\
$^{24}$\footnotesize{\textit{INAF, Osservatorio Astronomico di Trieste, via G. B. Tiepolo 11, I-34143 Trieste, Italy}} \\
$^{25}$\footnotesize{\textit{Universit\'{a} di Salerno, Dipartimento di Fisica "E.R. Caianiello", Via Giovanni Paolo II 132, I-84084 Fisciano (SA), Italy}} \\
$^{26}$\footnotesize{\textit{INAF, Osservatorio Astronomico di Capodimonte, Via Moiariello 16, I-80131 Napoli, Italy}} \\
$^{27}$\footnotesize{\textit{INFN, Gruppo Collegato di Salerno - Sezione di Napoli, Dipartimento di Fisica "E.R. Caianiello", Universit\`{a} di Salerno, via Giovanni Paolo II, 132 - I-84084 Fisciano (SA), Italy}} \\
$^{28}$\footnotesize{\textit{Institut de Ci\`{e}ncies del Cosmos (ICCUB), Universitat de Barcelona (UB), Mart\'{i} i Franqu\`{e}s 1, E-08028 Barcelona, Spain}} \\
$^{29}$\footnotesize{\textit{Dribia Data Research S.L., Pg. de Gr\`{a}cia, 55, 3r 4a, 08007 Barcelona, Spain}} \\
$^{30}$\footnotesize{\textit{Aix Marseille Univ, CNRS, CNES, LAM, Marseille, France}} \\
$^{31}$\footnotesize{\textit{Institut d'Estudis Espacials de Catalunya (IEEC), Esteve Terradas, 1, Edifici RDIT, Campus PMT-UPC, 08860 Castelldefels (Barcelona), Spain}} \\
$^{32}$\footnotesize{\textit{Leiden Observatory, Leiden University, PO Box 9513, NL-2300 RA Leiden, The Netherlands}} \\
$^{33}$\footnotesize{\textit{NOVA Optical and Infrared Instrumentation Group at ASTRON, Oude Hoogeveensedijk 4, 7991 PD, Dwingeloo, The Netherlands}}\\
$^{34}$\footnotesize{\textit{School of Physical and Chemical Sciences - Te Kura Mat\={u}, University of Canterbury, Private Bag 4800, Christchurch 8140, New Zealand}} \\
$^{35}$\footnote{\textit{RALSpace, STFC, Harwell, Didcot OX11 0QX, UK.}}

\section{Additional material}\label{appendix2}
\begin{figure*}
    \centering
    \includegraphics[width=1.\textwidth]{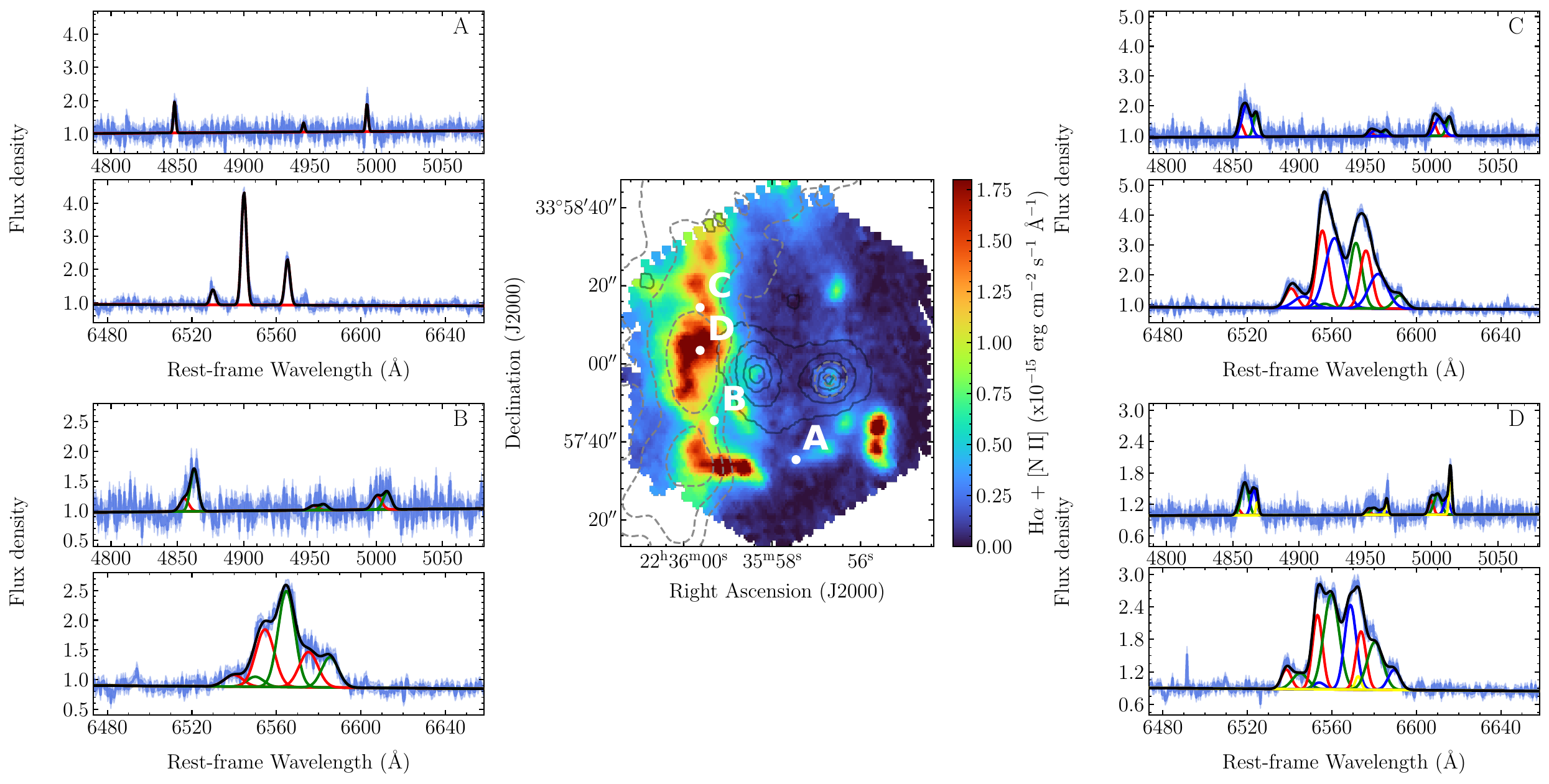}
    \caption{Example fits in the rest-frame of the system of the four Gaussian models at different locations in Stephan's Quintet as denoted in by the letters in the H$_{\alpha}$ map in the middle panel (the same as in Figure \ref{fig:vel_map}). The flux density $\pm1\sigma$ is indicated in blue, whereas the best-fit model is presented in black. The separate Gaussian components are also shown in various colours.}
    \label{fig:example_fits}
\end{figure*}

\begin{figure*}
    \centering
    \includegraphics[width=.7\textwidth]{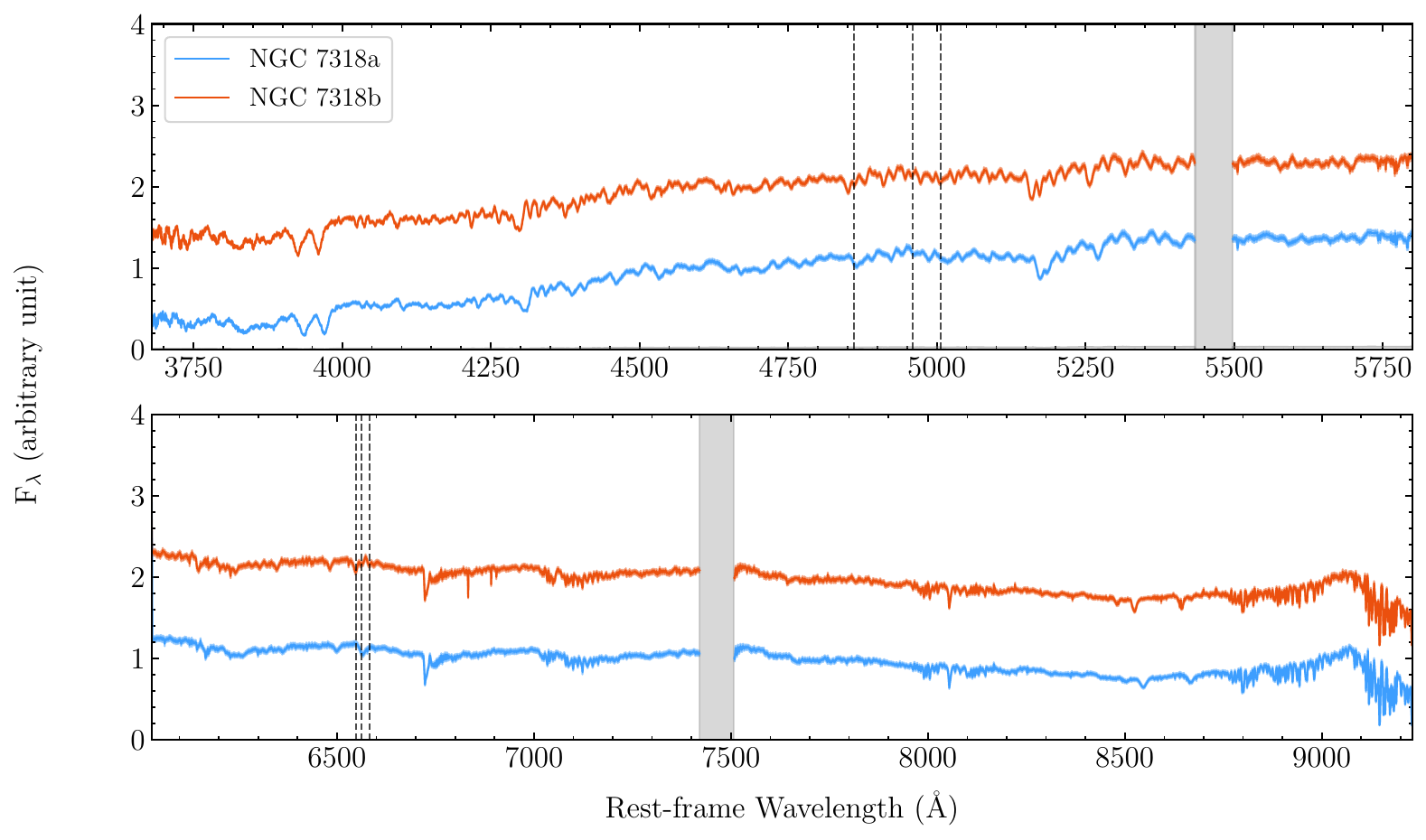}
    \caption{The spectra of NGC 7318a$\&$b from the WEAVE First Light data, created by summing all spaxels inside a circular aperture with a 5 arcsec radius centred on each galaxy. The spectra have been arbitrarily normalised and offset from one another for comparison. The grey shaded spectral regions denote the chip gaps, while the black dotted lines show the central rest-frame wavelength of the H$\beta\,\lambda$4861, [\textsc{Oiii}]\,$\lambda\lambda$4959\slash 5007, H$\alpha$\,$\lambda$6563, and [\textsc{Nii}]\,$\lambda\lambda$6548,6583 emission lines.}
    \label{fig:gal_spec}
\end{figure*}

\begin{figure*}
    \centering
    \includegraphics[width=1.\textwidth]{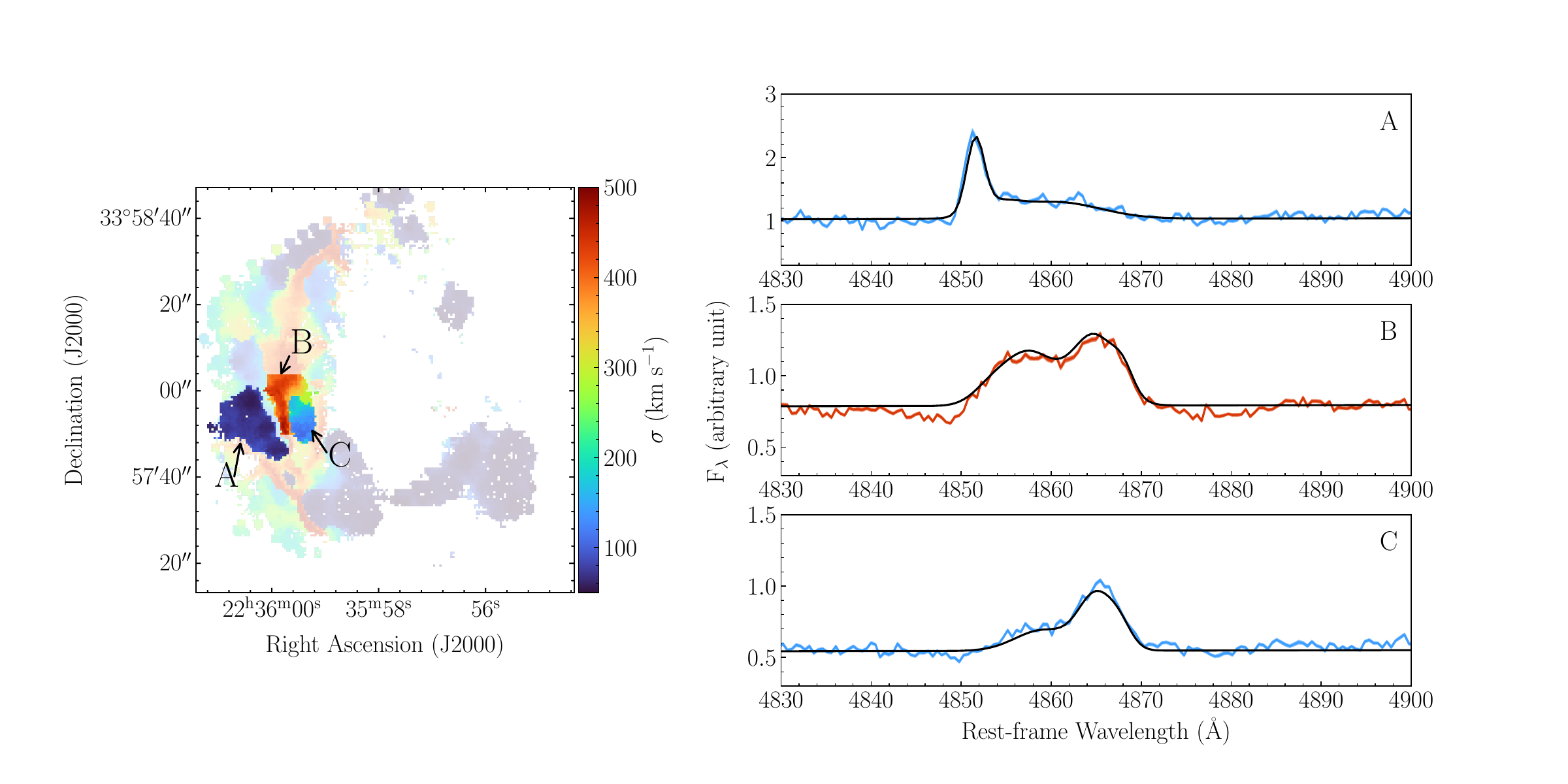}
    \caption{Line profiles in different regions of the LSSR around SQ. The left panel shows the velocity dispersion of H$\alpha$ as in Figure \ref{fig:Halpha}, along with three regions of interest as indicated by the arrows. The right panel presents the stacked spectra of these regions around the H$\beta$ line (chosen since it is well isolated from other species, unlike e.g. H$\alpha$), where the shaded area indicates the $\pm1 \sigma$ uncertainty on the stack. The black solid line shows the best fit to the data, where the emission lines in all three stacked spectra has been modelled by four Gaussian components, as determined by the BIC.}
    \label{fig:reg_stacks}
\end{figure*}

\begin{figure*}
    \centering
    \includegraphics[width=1\textwidth]{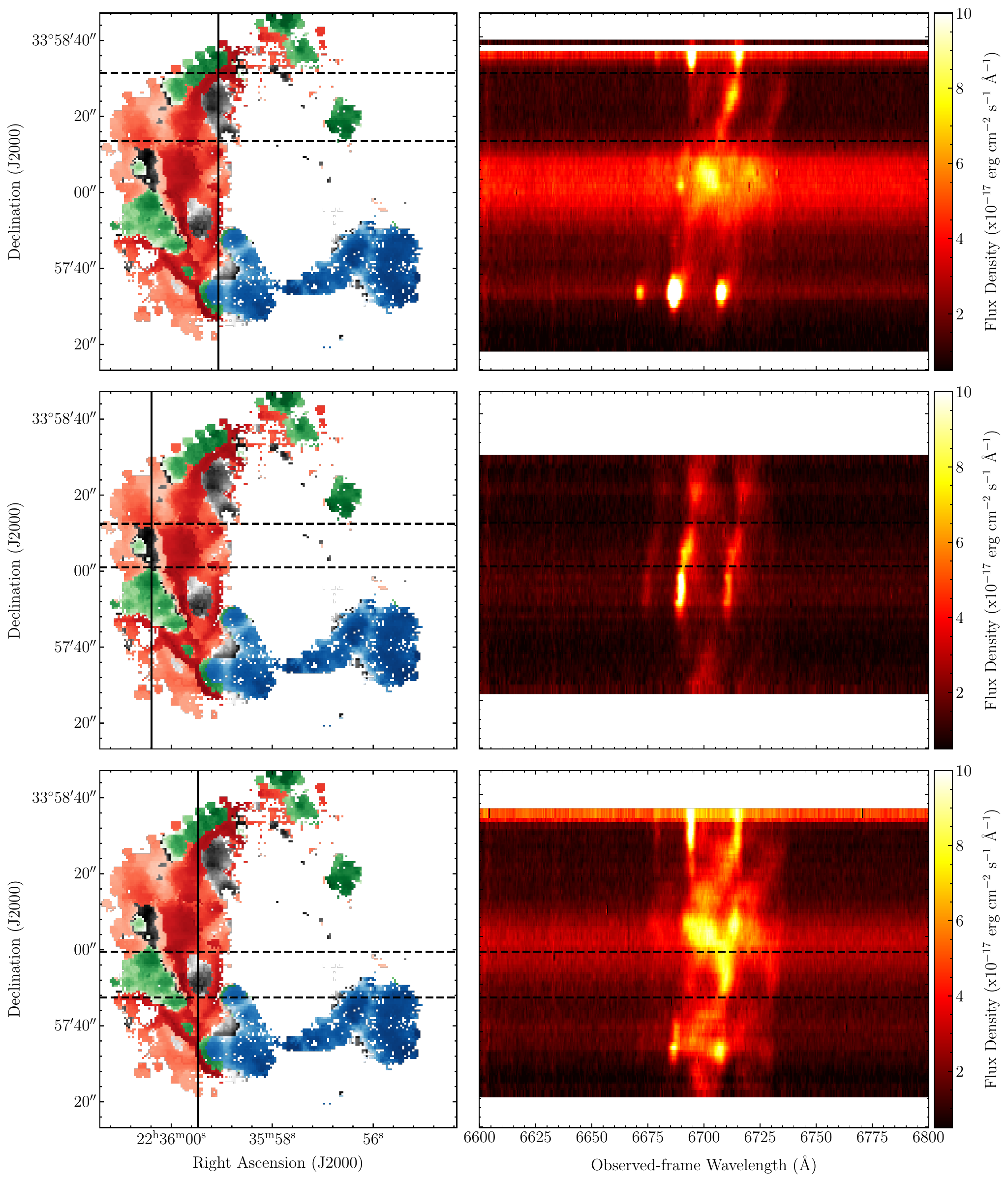}
    \caption{Position-velocity diagrams, similar to Figure \ref{fig:vel_map}, but here centered on the bubbles. The left panels present the different kinematically-defined regions as in Figure \ref{fig:v_sigma}, where the solid lines denote the 2D spectrum shown in the right panels, and the dashed lines serve as guides to trace the location of the different bubbles.}
    \label{fig:vel_map_bubbles}
\end{figure*}

In this appendix, we provide additional material:
\begin{itemize}
    \item Figure \ref{fig:example_fits} contains examples of how our line fitting technique -- which is described in section \ref{sec:method} -- uses up to four Gaussian components per emission line species to disentangle the complex velocity structure in SQ. We have identified four different example spaxels, located across the field of view (denoted in the middle panel), each modelled with different number of Gaussian components, with the number required determined using the BIC. The outer panels (labelled A, B, C \& D for spaxels which require 1, 2, 3 \&\ 4 velocity components per species, respectively) show the wavelengths around H$\beta$ and [\textsc{Oiii}]\,$\lambda\lambda$4959\slash 5007 lines in the upper axis, and the H$\alpha$\,$\lambda$6563 plus [\textsc{Nii}]\,$\lambda\lambda$6548,6583 complex on the lower axis. 
    
    \item Figure \ref{fig:gal_spec} shows stacked spectra of NGC 7318 a\&b (in blue and red, respectively), with the absence of emission lines abundantly clear. We have included stacked spectra of the two galaxies in both the blue and red arms of the WEAVE spectrograph, and have denoted the rest-frame wavelengths of the prominent emission lines used in this study, namely the H$\beta\,\lambda$4861, [\textsc{Oiii}]\,$\lambda\lambda$4959\slash 5007, H$\alpha$\,$\lambda$6563, and [\textsc{Nii}]\,$\lambda\lambda$6548,6583 emission lines.
    
    \item Figure \ref{fig:reg_stacks} compares the average velocity profiles of some example features dynamically identified in section \ref{sec:weave_overview}, showing three regions (labelled A, B \&\ C) which are neighbouring in projection (left panel) have very different emission line profiles (right panels, showing the well isolated H$\beta$ line). It is clear that these differences are not artefacts of the method used to calculate the effective velocity dispersion ($\sigma$), where we take the FWHM of the total line complex, irrespective of its shape, and convert it to $\sigma$ assuming that $\sigma = \mathrm{FWHM} \slash 2.355$ as in the limit of Gaussian line profiles, even though e.g. Figure \ref{fig:example_fits} highlights that in many cases the profiles are complex. We have also included our best-fit model for each spectrum as a solid black line, which demonstrates the once more the robustness of the spectral fitting method described in section \ref{sec:method} in handling complex emission line profiles.

    \item Figure \ref{fig:vel_map_bubbles} contains three further position-velocity diagrams -- similar in layout to Figure \ref{fig:vel_map} in section \ref{sec:data_weave}. In these figures, the 2D spectra visible in the right hand panels correspond to slices of spaxels chosen to coincide with the dynamically-defined `bubbles' (shaded grey in the left-hand panels). Their approximate vertical extent is indicated by dashed horizontal lines crossing the full row to enable the reader to compare the velocity location and profile of the `bubble' spaxels with those of their neighbours (which can be dynamically-identified as `shock' or \textsc{Hii} regions). In all cases it is clear that so-called `bubbles' have velocity structures that significantly differ from the spaxels closest in projection, and that their identification is therefore not the result of e.g. chance projection of unrelated structure along the line of sight, but likely a physical property of the system.  

\end{itemize}

\bsp	
\label{lastpage}
\end{document}